\documentclass[prd,preprint,aps,amsmath,amssymb,showpacs,nofootinbib]{revtex4-1}
\usepackage{graphicx}
\usepackage{enumitem}
\usepackage{slashed}
\usepackage{url}
\usepackage[T1]{fontenc}
\usepackage{bm}
\usepackage{color}
\usepackage[export]{adjustbox}
\usepackage{ulem}
\usepackage{hyperref}
\hypersetup{
    colorlinks=true,
    linkcolor=red,
    filecolor=cyan,      
    urlcolor=magenta,
    citecolor=blue
    }

\newcommand{\GeV}{\rm GeV}

\renewcommand\[{\begin{equation}}
\renewcommand\]{\end{equation}}

\begin{document}

\preprint{PITT-PACC-2213}

\title{Dynamics of Dark Matter Misalignment Through the Higgs Portal}
\author{Brian Batell}
\email{batell@pitt.edu}
\affiliation{Pittsburgh Particle Physics, Astrophysics, and Cosmology Center, Department of Physics and Astronomy, University of Pittsburgh, Pittsburgh, USA}
\author{Akshay Ghalsasi}
\email{akg53@pitt.edu}
\affiliation{Pittsburgh Particle Physics, Astrophysics, and Cosmology Center, Department of Physics and Astronomy, University of Pittsburgh, Pittsburgh, USA}
\author{Mudit Rai}
\email{mur4@pitt.edu}
\affiliation{Pittsburgh Particle Physics, Astrophysics, and Cosmology Center, Department of Physics and Astronomy, University of Pittsburgh, Pittsburgh, USA}


\begin{abstract}
  A light singlet scalar field feebly coupled through the super-renormalizable Higgs portal provides a minimal and well-motivated realization of ultra-light bosonic dark matter. We study the cosmological production of dark matter in this model by elucidating the dynamics of two sources of scalar field misalignment generated during the radiation era. For large scalar masses (above ${\cal O}(10^{-3}\,{\rm eV}$)), dark matter is produced through {\it thermal misalignment}, by which the scalar field is driven towards large field values as a result of the finite-temperature effective potential. The dominance of thermal misalignment in this mass range leads to a sharp relic abundance prediction which is, to a significant extent, insensitive to the initial conditions of the scalar field. 
  On the other hand, for low mass scalars (below ${\cal O}(10^{-5}\,{\rm eV}$)), dark matter is produced via {\it VEV misalignment}, which is caused by the induced scalar field vacuum expectation value triggered by the electroweak phase transition. We show that the relic abundance in this low mass range is sensitive to the scalar field initial conditions. In the intermediate mass range, the relic abundance is a consequence of a competition between thermal misalignment and VEV misalignment, which can potentially lead to novel forced resonance effects which cause a recurring enhancement and suppression in the late time oscillation amplitude as a function of the scalar mass. 
  We compare our relic abundance predictions with constraints and projections from equivalence principle and inverse square law tests, stellar cooling, resonant molecular absorption, and observations of extra-galactic background light and diffuse X-ray backgrounds. New experimental ideas are needed to probe most of the cosmologically motivated regions of parameter space.
\end{abstract}

\maketitle

\section{Introduction}

Understanding the nature of the cosmic dark matter (DM), which constitutes roughly a quarter of the energy density in the universe~\cite{Planck:2018vyg}, is among the most pressing problems in particle physics and cosmology today. Despite its clear influence on a host of astrophysical and cosmological phenomena, the most basic properties of DM remain mysterious, including it fundamental dynamics (spin, mass, interactions, etc.) as well as its origin during the earliest epochs of the universe.

One particularly compelling idea is that DM is an ultra-light, feebly-coupled scalar field $\phi$~\cite{Antypas:2022asj}. As is well-known, such scalar field DM is generically created in the early universe through the misalignment mechanism~\cite{Preskill:1982cy, Abbott:1982af, Dine:1982ah}. 
The scalar field starts from some initial field value at the end of inflation that is misaligned with respect to its eventual potential minimum (i.e., its vacuum expectation value (VEV)).
In the early stages of the radiation-dominated era, the scalar field is held up by Hubble friction and, notably, does not experience any additional dynamical misalignment during this epoch. 
As the universe expands, the scalar field eventually begins oscillating once the Hubble expansion rate drops below its mass. 
The oscillating scalar field subsequently forms a pressureless, non-relativistic fluid and thus has the desired bulk properties of cold DM.
Thus, in this {\it standard misalignment} scenario, the DM density today is controlled by the initial value $\phi_{i}$ and does not depend at all on the coupling of $\phi$ to the Standard Model (SM).

A simple model realization of ultra-light bosonic DM consists of a real singlet scalar field coupled to the SM through the super-renormalizable Higgs portal, 
$A\phi H^\dag H$, as first proposed more than a decade ago by Piazza  and Pospelov~\cite{Piazza:2010ye}. This model is attractive from several perspectives.  First, it is among the most minimal, UV-complete extensions of the SM, with the cosmology (up to initial conditions) and phenomenology determined by two parameters, namely, the mass of the scalar, $m_\phi$ and its Higgs portal coupling $A$. 
Second, the feebleness of this coupling allows the light scalar mass (and scalar potential) to be stable against radiative corrections. 
Furthermore, the model has a distinctive phenomenology, with a variety of existing probes from both terrestrial experiments and astrophysical observations, and has served as a well-motivated benchmark model for several proposed ultralight scalar DM detection concepts; see, e.g., Refs.~\cite{Graham:2015ifn,Arvanitaki:2017nhi,Antypas:2022asj}.

On the other hand, the cosmology of this simple model remains relatively underexplored, and it is the primary aim of this work to fill that gap. As our study will make clear, the cosmology of the scalar field in the Higgs portal model is not generally encapsulated by the standard misalignment scenario discussed above. In contrast to the standard misalignment scenario, the scalar field  undergoes a nontrivial evolution during the radiation era by which new sources of misalignment are dynamically generated, impacting the scalar relic abundance in an essential way.
In particular, we will investigate two distinct dynamical misalignment mechanisms that are inherent in the Higgs portal model: (i) {\it thermal misalignment}, and (ii) {\it VEV misalignment}.

The thermal misalignment mechanism was recently explored as a generic DM production mechanism in Ref.~\cite{Batell:2021ofv} (see also Refs.~\cite{Buchmuller:2003is,Fardon:2003eh,Buchmuller:2004xr,Moroi:2013tea,Lillard:2018zts,Chun:2021uwr,Brzeminski:2020uhm,Croon:2022gwq,Cheek:2022yof} for related work) in the context of simple Yukawa-type theory, but the essential aspects of the mechanism carry over to the Higgs portal model. 
In the earliest stages of the radiation era, the scalar field responds to a finite-temperature correction to its effective potential, associated with the free energy density of the Higgs degrees of freedom in the thermal bath to which the scalar is feebly coupled. This tends to drive the scalar field toward its high-temperature minimum at large field values, thus generating misalignment. As we will demonstrate below, thermal misalignment dominates for scalar masses larger than a few meV. Furthermore, provided the initial field value $\phi_i$ at the end of inflation is small (in magnitude) in comparison to the displacement generated by thermal misalignment, the late-time scalar oscillation amplitude and associated relic density  are insensitive to initial conditions, instead being tightly controlled by the DM mass and Higgs portal coupling. Thus, a sharp prediction can be made for the model parameters (scalar mass and Higgs portal coupling) yielding  the observed relic abundance through thermal misalignment, providing a cosmologically motivated target that can be compared with experimental tests. 

VEV misalignment is a second source of dynamical misalignment which is built in to the Higgs portal model. It arises as a consequence of the electroweak phase transition (EWPT), during which the Higgs field quickly turns on as electroweak symmetry is broken and induces a rapid shift in the $\phi$ VEV towards its zero-temperature VEV. As we will see, VEV misalignment dominates for scalar masses below about $10^{-5}$ eV. For this region of parameter space, we will find that the scalar oscillation amplitude and corresponding relic abundance sensitively depends on the initial field value $\phi_i$, in particular, whether the scalar field is initially close to its zero temperature VEV or far away from it. 

In the intermediate scalar mass range, an intricate interplay between thermal misalignment and VEV misalignment may potentially give rise to a novel forced resonance effect, with the scalar oscillation amplitude experiencing either an enhancement or suppression following the EWPT depending on the $\phi$ mass. This manifests as a striking series of peaks and valleys in the relic density target line as the scalar mass is varied. 

\begin{figure}[h!]
\includegraphics[width=\textwidth, center]{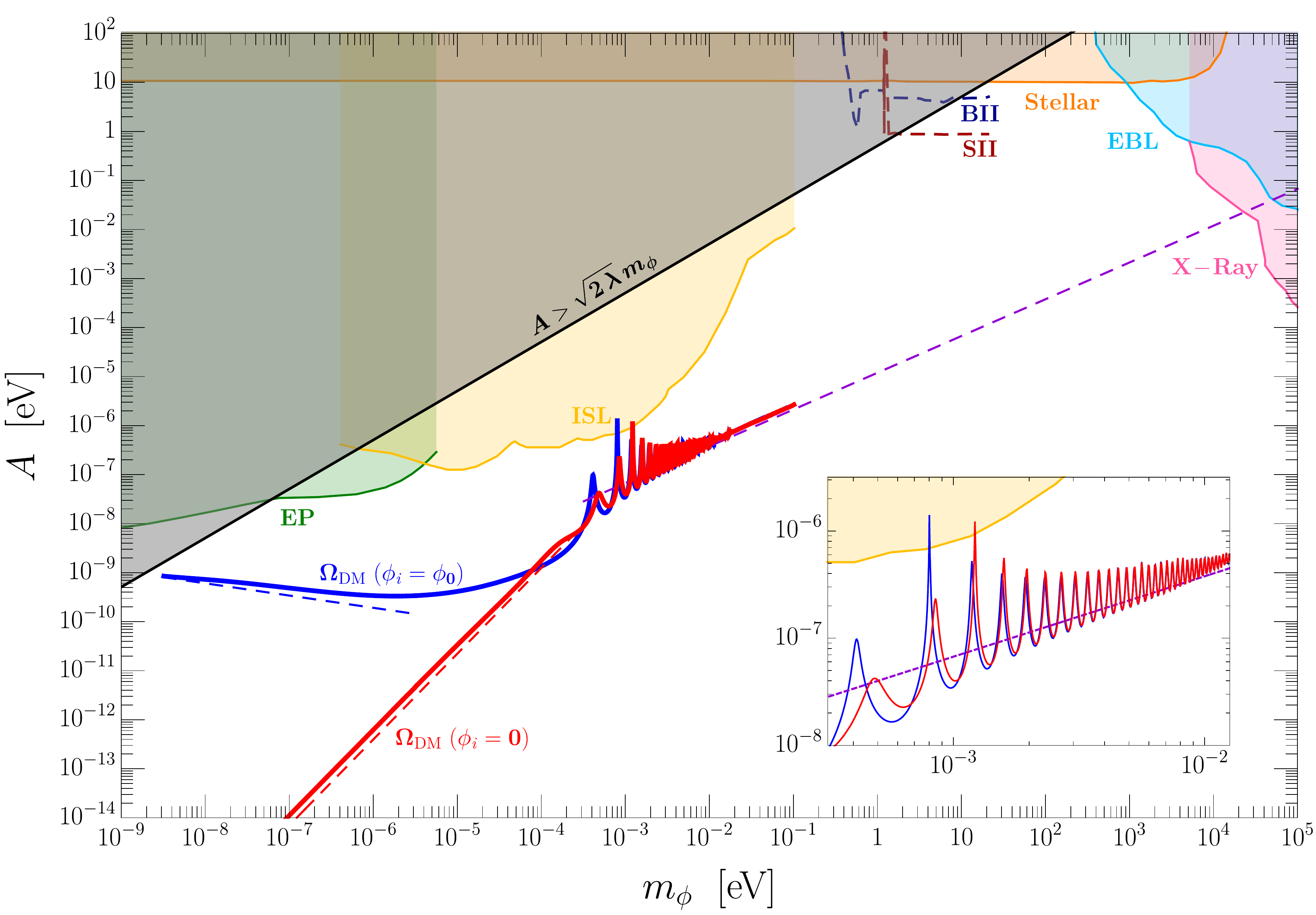}
\caption{ 
DM relic abundance predictions along with constraints and projections 
displayed in the $m_\phi - A$  (mass-coupling) plane. 
Parameter choices leading to the observed relic abundance,  $\Omega_{\phi,0} = \Omega_{\rm DM} = 0.26$, computed by numerically evolving Eq.~(\ref{eq:mastereq}) and using Eq.~(\ref{eq:phi-relic-density}), are shown for two choices of initial conditions: $\phi_i = \phi_0$ (blue solid) and $\phi_i = 0$ (red solid). 
In Region I, thermal misalignment dominates and the approximate relic abundance is given by Eq.~(\ref{eq:ODMreg1}) (purple dashed), independently of the choice of initial conditions. 
In Region III, VEV misalignment dominates over thermal misalignment, and the relic abundance is sensitive to the initial conditions, being given by Eq.~(\ref{eq:ODMreg3}) (blue dashed) for $\phi_i = \phi_0$ and by Eq.~(\ref{eq:ODMreg31}) (red dashed) for $\phi_i = 0$.
In the intermediate mass range the relic density lines feature a series of peaks and valleys, the result of a forced resonance phenomena due to the EWPT (see inset plot).
Also shown are bounds and projections from equivalence principle tests (EP, green), inverse square law tests (ISL, yellow), stellar cooling (orange), resonant absorption in molecules [SII (dark red dashed) and BII (dark blue dashed)], extra-galactic background light (EBL, light blue), and X-ray observatories (X-Ray, pink), and a viable electroweak vacuum (\ref{eq:stability}) (gray). See the text for further details.
}
\label{fig:relic}
\end{figure}

Our main results are presented in Fig.~\ref{fig:relic} as contours in the mass-coupling parameter space yielding  the observed DM relic abundance,  providing cosmologically motivated targets for experiments searching for ultralight scalar DM. 
In view of this, we also compile the existing bounds and sensitivity projections for a variety of experimental and astrophysical probes of the model, including equivalence principle and inverse square law tests, stellar cooling, resonant molecular absorption, and observations of extra-galactic background light and diffuse X-ray backgrounds. 
Still, much of the parameter space remains unexplored, and we hope our results will stimulate new innovative experimental concepts for ultra-light DM searches. 

It is important to note that the potential impact of the finite-temperature effects and the EWPT on the scalar field evolution and relic abundance were discussed in the original study of Ref.~\cite{Piazza:2010ye}. The effect of the EWPT on scalar field misalignment in Higgs portal models was also investigated in Ref.~\cite{Arkani-Hamed:2020yna}, though thermal effects were not considered in that work. 
We believe our study elucidates the dynamics of the scalar field during the radiation era by including both the thermal misalignment and VEV misalignment effects, discerning the regions of parameter space where each is relevant, and exploring the role of and sensitivity to initial conditions. 
In particular, we have carefully derived estimates of the scalar relic density, both by numerically solving its equation of motion during the radiation era through the EWPT and by developing various approximate solutions as appropriate for the region of parameter space, cosmological epoch, and initial conditions under consideration. Both our numerical and analytical results are shown in Fig.~\ref{fig:relic} and are in good agreement. 

We emphasize that our study focuses on a restricted form of the scalar potential which, along wth the usual Higgs potential terms, includes only a scalar mass term and the super-renormalizable Higgs-portal coupling term $A \phi |H|^2$. This is clearly not the most general $\phi$ scalar potential. If additional large scalar interactions were present, e.g., large linear $\phi$, cubic $\phi^3$, or quartic $\phi^4$ terms, such terms could significantly impact our results and would require further study going beyond our present scope. On the other hand, our assumed starting point with the restricted form of the scalar potential can be considered to be technically natural due to the approximate shift symmetry of the scalar field that is only broken by the feeble Higgs portal coupling $A \ll m_\phi$ and scalar mass. Thus, at least within our low energy effective theory, the radiatively generated potential terms will have negligible impact on our results.

This paper is organized as follows. In Sec.~\ref{sec:model} we describe the super-renormalizable Higgs portal model, including its zero temperature vacuum and scalar spectrum. 
Sec.~\ref{sec:cosmology} describes the cosmology of the scalar field and forms the bulk of the paper.
It contains several subsections covering the effective potential, EWPT, sources of misalignment and initial conditions, scalar field equation of motion and evolution, and analytic relic density estimates.
Sec.~\ref{sec:results} compares our relic abundance predictions with existing and proposed probes of the model. 
Finally, our conclusions are presented in Sec.\ref{sec:conclusion}.

\section{Super-Renormalizable Higgs Portal Model}
\label{sec:model}

We consider the model of Ref.~\cite{Piazza:2010ye} which contains a real singlet scalar field $\phi$ that couples to the SM via the super-renormalizable Higgs portal. The scalar potential to be considered is
\begin{equation}
    V =-\mu^2 \, H^\dag H + \lambda \, (H^\dag H)^2 + \frac{1}{2}  m_\phi^2 \, \phi^2 + A \, \phi \, H^\dag H\,,
    \label{eq:potential}
\end{equation}
where the first two terms constitute the usual SM Higgs potential, $m_\phi$ is the scalar mass parameter, and $A$ is a dimensionful coupling of the scalar to the Higgs. Note that we have omitted a potential linear term in $\phi$, which can always be achieved by performing a field redefinition. 

We will mainly be interested in feeble couplings such that $A \ll m_\phi$. This implies that the light scalar mass is technically natural, with loops generating $\delta m_\phi^2 \sim A^2/(16 \pi^2) \log{\Lambda_{\rm UV}} \ll m_\phi$. This is one of the attractive features of the super-renormalizable portal. 
Furthermore, radiatively generated nonlinear potential terms such as $\phi^3$ and $\phi^4$ are also small and will be neglected for the remainder of this work. 

\subsection{Zero temperature vacuum and spectrum}

We first consider the theory at zero temperature. The background Higgs field is parameterized as $H^T = (0, \tfrac{1}{\sqrt{2}}h)$.
The potential then reads
\begin{align}
V_0(\phi, h) = -\frac{1}{2}\,\mu^2 \, h^2 + \frac{1}{4}\lambda \, h^4 + \frac{1}{2} m_\phi^2 \phi^2 + \frac{1}{2} A \, \phi \, h^2\,.
\label{eq:potential1}
\end{align}
Minimizing the potential, we obtain the vacuum expectation values (VEVs) for the scalar fields $\langle h \rangle = v$ and $\langle \phi \rangle = \phi_0$, given by
\begin{align}
\label{eq:vphi0}
v^2 = \frac{\mu^2}{\lambda - A^2/2 m_\phi^2}\,, ~~~~~~~ \phi_0 = - \frac{A v^2}{2 m_\phi^2}\,,
\end{align}
with $v = 246$ GeV. It can be seen from Eq.~(\ref{eq:vphi0}) that a viable electroweak vacuum is obtained only for 
\begin{align}
    \label{eq:stability}
    \frac{A^2}{2 \, \lambda \, m_\phi^2} < 1\,.
\end{align}

To study the spectrum at zero temperature, we replace $h \rightarrow v +  \tilde{h}$ and $\phi \rightarrow \phi_0 + \tilde{\phi}$ in the potential, Eq.~(\ref{eq:potential1}), where $\tilde{h}$ and $\tilde{\phi}$ represent the fluctuations about the vacuum. Expanding the potential to quadratic order, we find that there is mass mixing between the scalars, 
\begin{align}
V & \supset \frac{1}{2} (2 \lambda v^2) \tilde{h}^2 + \frac{1}{2} m_\phi^2 \,\tilde{\phi}^2 + A \, v \, \tilde{h} \,\tilde{\phi}\,.  
\end{align}
We move to the physical basis by performing a rotation, 
\begin{align}
\left[
\begin{array}{c}
\tilde{h} \\
\tilde{\phi}
\end{array}
 \right] 
 =
 \left[
\begin{array}{cc}
\cos \theta & \sin \theta \\
-\sin\theta & \cos\theta
 \end{array}
 \right]
 \left[
\begin{array}{c}
 h \\
 \phi
\end{array}
 \right],
 \label{eq:rotation}
\end{align}
where, in an abuse of notation, $h$ and $\phi$ represent the physical scalar fluctuations in the mass basis.
The mixing angle $\theta$ in (\ref{eq:rotation}) is given by 
\begin{equation}
\tan 2 \theta = \frac{2 A v}{2 \lambda v^2 - m_\phi^2}\,.
\end{equation}
The mass eigenvalues are 
\begin{equation}
M^2_{h,\phi} = \frac{1}{2} \left[ 2 \lambda v^2 + m_\phi^2 \pm \sqrt{  (2 \lambda v^2 - m_\phi^2)^2 + 4 A^2 v^2 }    \right]\,.
\end{equation}
As we will always be working in the regime $A^2 \leq 2 \lambda m_\phi^2 \ll \lambda v^2$, the approximate expressions for the mixing angle and mass eigenvalues are
\begin{align}
\theta \sim \frac{A}{2\lambda v} \simeq \frac{A v}{M_h^2}\,, ~~~~~~
M_h^2 \simeq 2 \lambda v^2 + \frac{A^2}{2 \lambda}\,,~~~~~~M_\phi^2 \simeq m_\phi^2 -  \frac{A^2}{2 \lambda}\,.
\label{eq:angle-mass}
\end{align}
Note that the correction to the Higgs mass is always negligible and thus $M_h^2 \simeq 2 \lambda v^2 \equiv m_h^2 = (125~\GeV)^{2}$ as in the SM. 
Furthermore, we will find that in most of the cosmologically interesting parameter space, $A \ll m_\phi$, such that $M^{2}_{\phi} \simeq m^{2}_{\phi}$.
Note that very close to the boundary where the electroweak vacuum is viable (see Eq.~(\ref{eq:stability})), $A^2/(2 \lambda m_\phi^2) \lesssim 1$, and only in this very small region of parameter space is the physical scalar mass is substantially modified from $m_\phi^2$. 

The electroweak vacuum condition, Eq.~(\ref{eq:stability}) (gray shaded region in Fig.~\ref{fig:relic}), could be relaxed in a scenario with additional scalar potential terms. However, in this case the light scalar mass $M_\phi$ would require fine-tuning, as can be seen from Eq.~(\ref{eq:angle-mass}) above. In this work, we will restrict our investigation of the scalar cosmology to parameters satisfying Eq.~(\ref{eq:stability}).

\section{Cosmology}
\label{sec:cosmology}
In this section, we describe the cosmological evolution of the scalar $\phi$ before and through the EWPT until it oscillates about its late time potential minimum and behaves as DM. We will assume that the universe reheats to a temperature much larger than the electroweak scale $(T_{\rm RH} \gg v)$. We begin by describing the contribution to the scalar potential from the thermal bath. We will then describe the EWPT experienced by the Higgs, which can significantly impact the dynamics of $\phi$ in certain regions of parameter space. We describe the sources of misalignment and detail the initial conditions assumed for the scalar at the end of inflation. We then describe the final equation of motion that governs the scalar evolution and numerically estimate the relic $\phi$ abundance. We also derive approximate analytic estimates of the relic density and compare them with our numerical results.

\subsection{Effective Potential}
\label{sec:Veff}
Our analysis begins in the radiation-dominated era during which the Hubble parameter is given by $H = 1/2t = \gamma T^2/M_{\rm pl}$, where $t$ is the cosmic time, $T$ is the temperature of the SM thermal bath, $M_{\rm pl} = 2.43 \times 10^{18}$ GeV is the reduced Planck mass, and $\gamma(T) \equiv \sqrt{\pi^2 g_*(T)/90}$ with $g_{*(S)}$ the effective number of relativistic (entropy) degrees of freedom. During this epoch, the effective potential of the scalar fields is given by 
\begin{align}
\label{eq:Veff}
V_{\rm eff}(\phi,h,T)& =V_0(\phi,h) 
+ V_{\rm CW}(\phi,h)
+V_{T}(\phi,h,T)\,.
\end{align} 
The first piece is the tree-level potential given in Eq.~(\ref{eq:potential}). 
The second term represents the zero-temperature correction to the effective potential, i.e., the Coleman-Weinberg potential~\cite{Coleman:1973jx}, which we will comment on shortly. 
The final term in (\ref{eq:Veff}) is the finite-temperature correction associated with the thermal free energy density of the SM particles in the plasma~\cite{Dolan:1973qd,Weinberg:1974hy}. The expression for the finite-temperature potential in Landau gauge is~\cite{Quiros:1999jp}
\begin{align}
\label{eq:VT}
V_T(\phi,h,T) &
\supset \frac{1}{2\pi^2}T^4 J_B\left[ \frac{m_h^2(\phi,h,T)}{T^2} \right]
+\frac{3}{2\pi^2}T^4 J_B\left[ \frac{m_\chi^2(\phi,h,T)}{T^2} \right]
+ \frac{4}{2\pi^2}T^4 J_B\left[ \frac{m_{W_T}^2(h)}{T^2} \right] \nonumber  \\
& +  \frac{2}{2\pi^2}T^4 J_B\left[ \frac{m_{Z_T}^2(h)}{T^2} \right]
+ \frac{2}{2\pi^2}T^4 J_B\left[ \frac{m_{W_L}^2(h,T)}{T^2} \right]
+ \frac{1}{2\pi^2}T^4 J_B\left[ \frac{m_{Z_L}^2(h,T)}{T^2} \right]
\nonumber  \\
& +\frac{1}{2\pi^2}T^4 J_B\left[ \frac{m_{A_L}^2(h,T)}{T^2} \right] 
- \frac{12}{2\pi^2}T^4 J_F\left[ \frac{m_t^2(h)}{T^2} \right]\,, 
\end{align}
where 
the $J_{B,F}$ functions are defined as
\begin{align}
J_{B,F}(w^2) &= \int_0^\infty \! dx \,x^2 \, \log \bigg[1\mp \exp\left({-\sqrt{x^2+w^2}}\right)\bigg]\,.
\end{align}
The thermal squared masses, $m_i^2(\phi,h,T)$, entering in the arguments of these functions depend in general on both the background field fields $\phi$ and $h$ and the temperature. To obtain a reliable perturbative expansion near the EWPT, we follow the prescription for daisy resummation of Ref.~\cite{Parwani:1991gq}, writing the thermal squared mass as sum of a tree-level term field-dependent term, $m^2_{0,i}(\phi,h)$, and a temperature-dependent self-energy term $\Pi_i(T)$, 
\begin{align}
    m_i^2(\phi,h,T) = m^2_{0,i}(\phi,h) + \Pi_i(T)\,.
\end{align}
In particular, for the self-energies we retain the leading contributions in the high temperature expansion. 
The explicit expressions for the thermal masses are collected in App.~\ref{sec:thermal-masses}. 
The subscripts $(h,\chi,W_T,Z_T,W_L,Z_L,A_L,t)$ in these functions refer to the Higgs, Nambu-Goldstones, 
transverse and longitudinal gauge bosons, 
and top quark.
The prefactors in front of the $J_{B,F}$ functions account for the degrees of freedom of the corresponding field.

It is important to emphasize that $\phi$ itself is not in thermal equilibrium for the feeble values of Higgs portal coupling $A$ considered in this work, and as such there is no term associated with $\phi$ in the finite temperature potential~(\ref{eq:VT}). 
Rather, the finite temperature potential is due to the SM degrees of freedom in thermal equilibrium. It is a function of the background value for $\phi$ via the dependence of the squared mass parameters on $\phi$.

For simplicity, we will neglect the Coleman-Weinberg (CW) contribution $V_{\rm CW}(\phi,h)$ to the effective potential (\ref{eq:Veff}) in our numerical analysis below. The correction to the $\phi$ potential from $V_{\rm CW}$ is negligible in the viable region of parameter space (Eq.~(\ref{eq:stability})) due to the feeble coupling $A$, as already alluded to at the beginning of Sec.~\ref{sec:model}.
Furthermore, as we will see below, for large scalar masses the misalignment is generated at high temperatures (thermal misalignment) where $V_T(\phi,h,T)$ dominates and $V_{\rm CW}$ is negligible. For lower scalar masses, the misalignment is a result of the induced scalar VEV triggered by the EWPT. Even in this case,  $V_{\rm CW}$ does not qualitatively alter the behavior of the scalar potential and the nature of the EWPT, and will only lead to relatively small numerical differences in our results. 

Note that Eq.~(\ref{eq:VT}) is appropriate for temperatures above the QCD phase transition. For lower temperatures, an alternative description in terms of light hadronic states would be required. 
However, we will see the onset of scalar oscillations in the parameter space  we consider occurs at temperatures well above the QCD scale, so that a description at lower temperatures is not required to calculate the relic abundance. For similar reasons, the quarks lighter than the top quark as well as leptons will not influence our results.

We will find it convenient to work with the following set of dimensionless parameters: 
\begin{equation}
    \label{eq:dimless}
    y \equiv \frac{T}{\mu}\,,\quad~~ 
    \hat{\phi} \equiv \frac{\phi}{M_{\rm pl}}\,,\quad~~ 
    \hat{h} \equiv \frac{h}{\mu}\,, \quad~~
    \kappa \equiv \frac{m_{\phi} M_{\rm pl}}{\mu^{2}}\,,  \quad~~ 
    \beta \equiv \frac{A M_{\rm pl}}{\mu^2}\,.~~
\end{equation}
In particular, the parameter $y$ is a proxy for the temperature, with $y \sim {\cal O}(1)$ for temperatures near the electroweak scale. Furthermore, $\hat \phi$ and $\hat h$ are simply the ratios of the field variables to their respective characteristic scales. Finally, the scalar mass $m_\phi$ and Higgs portal coupling $A$ are expressed in terms of the dimensionless variables $\kappa$ and $\beta$, respectively.

Combining the $V_{0}$ and $V_{1}^{T}$ in Eqs.~(\ref{eq:potential},\ref{eq:VT}) and writing the effective potential in terms of dimensionless quantities ($\hat V_{\rm eff}=V_{\rm eff}/\mu^4$), we obtain
\begin{align}
\label{eq:Vhat}
\hat V_{\rm eff} & =-\frac{1}{2}\hat{h}^{2}(1-\beta \hat \phi)+\frac{1}{4}\lambda \hat{h}^{4}+\frac{1}{2}\kappa^{2} \hat{\phi}^{2}
  +\frac{y^{4}}{2\pi^{2}}\bigg\{
  J_{B}[\eta_{h}(\hat \phi, \hat h,\hat y)]
  + 3 J_{B}[\eta_{\chi}(\hat \phi, \hat h,\hat y)]
  + 4 J_{B}[\eta_{W_{T}}(\hat h)] 
  \nonumber \\ &  
  + 2 J_{B}[\eta_{Z_{T}}(\hat h)] 
  + 2 J_{B}[\eta_{W_{L}}(\hat h,\hat y)]
  + J_{B}[\eta_{Z_{L}}(\hat h,\hat y)]
  + J_{B}[\eta_{A_{L}}(\hat h,\hat y)] 
  - 12 J_{F}[\eta_{t}(\hat h,\hat y)]\bigg\}\,,
\end{align}
where we have defined the dimensionless arguments of the $J_{B,F}$ functions, $\eta_{i}(\hat \phi, \hat h, y) \equiv m^{2}_{i}(\phi,h,T)/T^{2}$. Explicit expressions for these quantities are provided in App.~\ref{sec:thermal-masses}.

\subsection{Higgs field and Electroweak Phase Transition}
\label{sec:EWPT}

We first discuss the evolution of the Higgs field $h$, which serves as the order parameter for the EWPT. In the SM, the EWPT is a smooth crossover~\cite{Kajantie:1996mn} characterized by the critical temperature $T_c \sim {\cal O}(v)$ ($y_c \equiv T_c /\mu \sim {\cal O}(1)$). The feeble portal coupling between $\phi$ and $h$ will not alter the nature of the phase transition, at least in the $\beta \hat \phi \ll 1$ regime of primary relevance for this work. 

We model the EWPT by assuming the Higgs field tracks its potential minimum derived from Eq.~(\ref{eq:Vhat}) throughout the phase transition, starting at $h = 0$ at high temperatures $(y \gg y_c)$, then taking nonzero values for $y < y_c$, and finally settling at  $h = v$ at low temperatures $(y \ll 1)$.  
While our treatment here has the benefit of simplicity, it would be very interesting to explore a more accurate modeling of the Higgs evolution using results from lattice studies of the EWPT in Refs.~\cite{Kajantie:1996mn,DOnofrio:2015gop}, which we leave to future work.

\begin{figure}
    \centering
    \includegraphics[width=0.6\columnwidth]{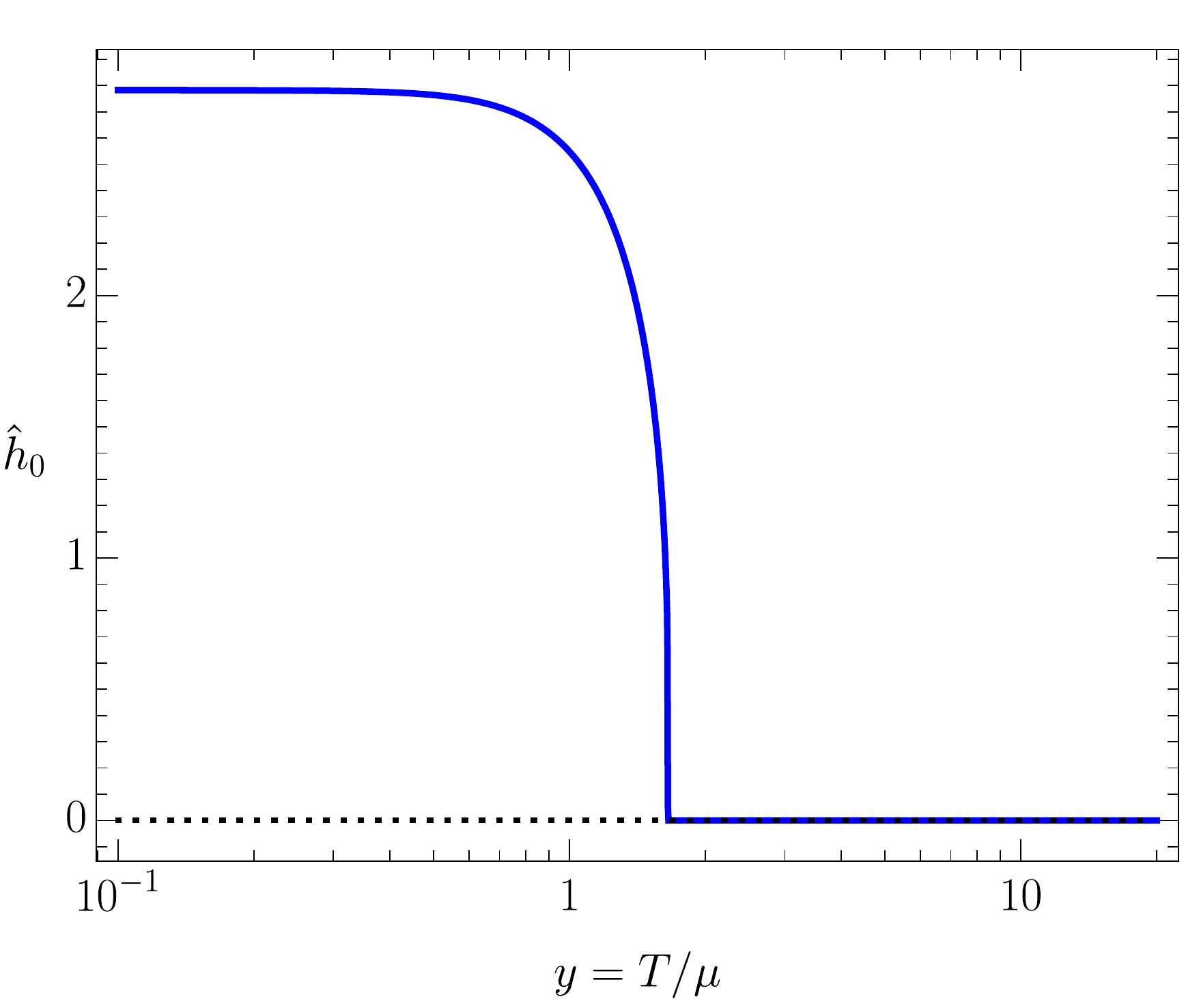}
    \caption{Dimensionless Higgs profile $\hat h_0 = \hat h(\hat \phi = 0,y)$ as a function of $y = T/\mu$.  
    The critical temperature corresponding to the EWPT can be seen at $y_{c} \simeq 1.6$ .
    }
    \label{fig:hath0}
\end{figure}

Thus, for temperatures below $y_{c}$, the evolution of $\hat{h}(\hat \phi,y)$ is determined by the minimization condition 
$\partial \hat V_{\rm eff}/\partial \hat h = 0$, with $\hat V_{\rm eff}$ given in Eq.~(\ref{eq:Vhat}):
\begin{align}
\label{eq:hath}
 0 &= \lambda \hat{h}^{2}-(1-\beta\hat{\phi})+\frac{y^{2}}{2\pi^{2}}
 \bigg\{
 6 \lambda (J'_{B}[\eta_{h}]+J'_{B}[\eta_{\chi}]) 
 + g^2 \left(2 J'_{B}[\eta_{W_{T}}] +J'_{B}[\eta_{W_{L}}]\right)  \nonumber \\
& + (g^{2}+g'^{2})J'_{B}[\eta_{Z_{T}}]
+ 2 y^2 \frac{\partial \eta_{Z_L}}{\partial \hat h^2}  J'_{B}[\eta_{Z_{L}}]
+   2 y^2 \frac{\partial \eta_{A_L}}{\partial \hat h^2}  J'_{B}[\eta_{A_{L}}]
-12 y_t^2 J'_{F}[\eta_{t}]
 \bigg\}\,.
\end{align}
Here $g$, $g'$, and $y_t$ are the $SU(2)_L$ gauge coupling, $U(1)_Y$ gauge coupling, and top Yukawa coupling, respectively, and 
the respective $\eta_{i}$ are given in App.~\ref{sec:thermal-masses}. 
Note that the $\eta_{i}$ are functions of $(\hat \phi, \hat{h},y)$. 
Our procedure is to solve Eq.~(\ref{eq:hath}) for $\hat{h}(\hat \phi,y)$ and then use this solution in equation of motion of $\hat{\phi}$, to be discussed below in Sec.~\ref{sec:scalar-dynamics}. 
In principle, Eq.~(\ref{eq:hath}) can be solved numerically.
However, since $\beta \hat{\phi} \ll 1$ for essentially all of our parameter space,
in practice we will treat $\beta \hat{\phi}$ as a perturbation. We therefore define $\hat{h}_{0}(y)$ as the solution to Eq.~(\ref{eq:hath}) with $\beta \hat{\phi} = 0$. Including the $\beta \hat \phi$ dependence will change the evolution and $y_{c}$ slightly 
($\propto \mathcal{O}(\beta \hat{\phi})$). The evolution of $\hat{h}_{0}(y)$ is shown in Fig.~\ref{fig:hath0}. 

\subsection{Sources of Misalignment and Initial Conditions}
\label{sec:ic}

\begin{figure}
    \centering
        \includegraphics[width=0.63\columnwidth]{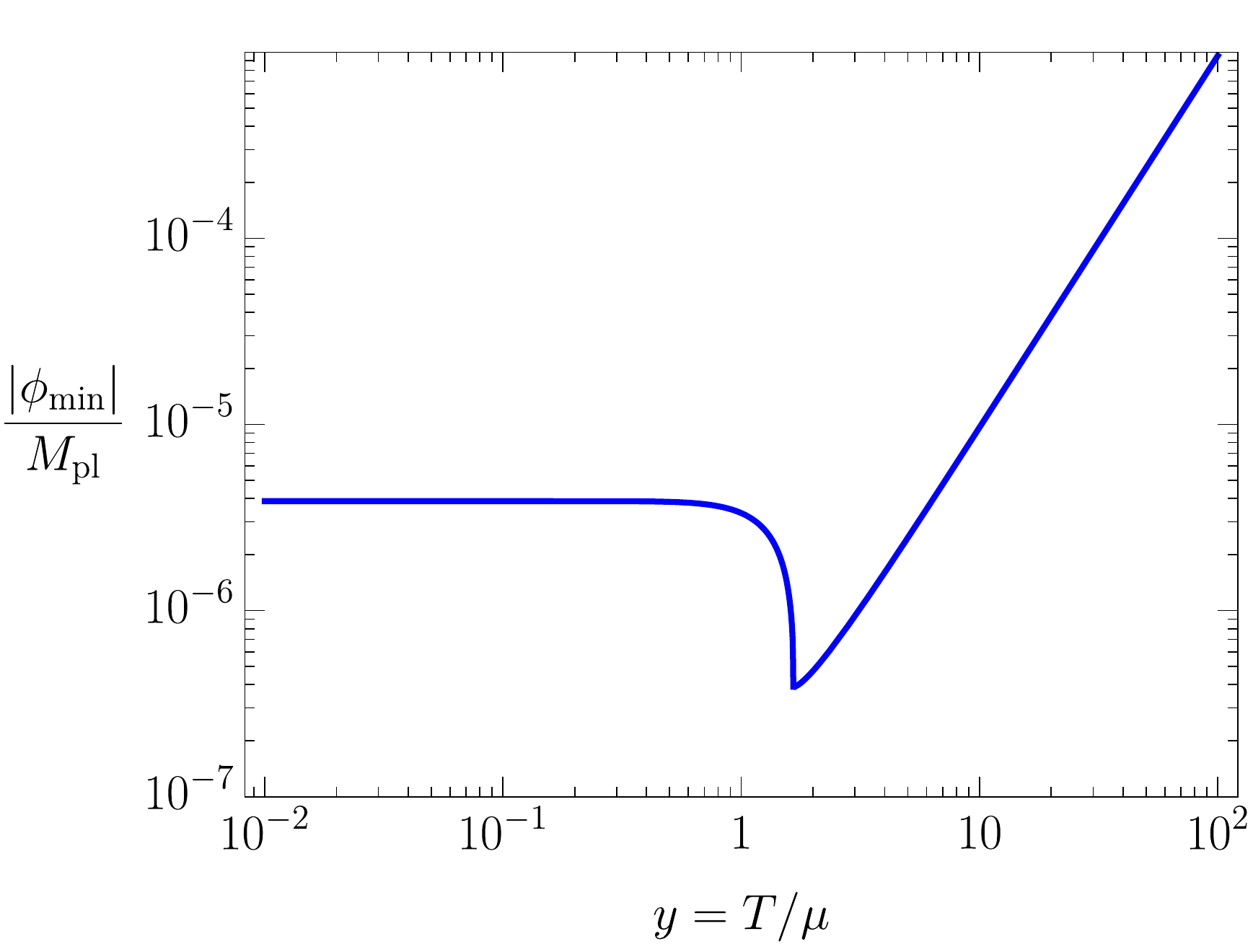}
\caption{Evolution of the temperature-dependent scalar field minimum  $|\hat \phi_{\rm min}(y) |$ as a function of $y  = T /\mu$ for the benchmark $\kappa =10^{3}$, $\beta = 1$. 
The $\phi$ minimum is controlled by the finite-temperature contribution to the effective potential for $y > y_{c}$, while instead for $y < y_{c}$ it transitions quickly towards its zero temperature value induced by the the Higgs VEV, Eq.~(\ref{eq:vphi0}). We have used $\hat h_0 = \hat h(\hat \phi = 0,y)$ as a function of $y = T/\mu$. 
}
    \label{fig:phi_min}
\end{figure}

Before we study the evolution of the scalar field $\phi$ in detail, it is worth discussing the various sources of misalignment that contribute to the late-time $\phi$ oscillation amplitude and associated relic density. 
To gain insight, it is helpful to examine the minimum of the effective potential with respect to $\hat \phi$ as a function of temperature, $\hat{\phi}_{\rm min}(y)$, which can be obtained from Eq.~(\ref{eq:Vhat}) by setting $\partial \hat{V}_{\rm eff}/\partial \hat{\phi} = 0$. 
The solution can be written as
\begin{align}
    \label{eq:phimin}
    \hat{\phi}_{\rm min} = -\frac{\beta}{2\kappa^{2}}\left[\hat{h}^{2} + \frac{y^{2}}{\pi^{2}}\left(J'_{B}[\eta_{h}]+3J'_{B}[\eta_{\chi}]\right)\right]\, .
\end{align}
Since $\beta \hat{\phi} \ll 1$ for the bulk of our parameter space, it is typically a good approximation to neglect this dependence in $\eta_{h,\chi}$ and $\hat h$ which enter in the r.h.s. of Eq.~(\ref{eq:phimin}). 
In Fig.~\ref{fig:phi_min} we show the variation of $\hat{\phi}_{\rm min}(y)$ with $y = T/\mu$ for a representative benchmark model. We see that at high temperatures, $y \gg y_c$, the minimum is located at large (negative)  scalar field values, reflecting the dominance of the second term in Eq.~(\ref{eq:phimin}). As the temperature drops, the minimum decreases as $|\hat{\phi}_{\rm min}| \propto y^2$ until the EWPT at $y_c$. At this point, the Higgs field turns on and the $\hat \phi$ VEV rapidly transitions to its zero temperature value, Eq.~(\ref{eq:vphi0}), or in terms of the dimensionless variables, 
\begin{equation}
\label{eq:phihat0}
 \hat {\phi}_{0} = -\frac{\beta}{  2 \lambda \kappa^{2} - \beta^{2} } \simeq -\frac{\beta}{  2 \lambda \kappa^{2} }\, .
 \end{equation}
The qualitative behavior of  $\hat{\phi}_{\rm min}$ is similar to that in Fig.~\ref{fig:phi_min} for other parameter choices. 

As we now discuss, there are three qualitatively distinct contributions to the misalignment of $\hat \phi$:
\begin{itemize}
\item \textbf{Misalignment due to initial conditions -}  This contribution is given by the value $\hat \phi_{i}$ the scalar field takes at the end of inflation.
\item \textbf{Thermal misalignment -} At temperatures much larger than the temperature of EWPT, $y \gg y_{c}$\,, the scalar experiences a finite-temperature contribution to the effective potential due to its feeble coupling to the thermalized Higgs degrees of freedom given by Eq.~(\ref{eq:VT}). This causes the scalar to roll towards the high-temperature minimum set by Eq.~(\ref{eq:phimin}) and illustrated 
in Fig.~\ref{fig:phi_min}, dynamically generating misalignment. 
We refer to this contribution as thermal misalignment, $\hat{\phi}_{T}$, and we will show below in Sec.~\ref{sec:Reg1} that
\begin{equation}
\label{eq:phiT}
\hat{\phi}_{T} \propto \frac{\beta}{\kappa} \,.
\end{equation}
\item \textbf{VEV misalignment -} Below the temperature of the EWPT, $y < y_{c}$\,, 
the Higgs field switches on and induces a rapid shift in the minimum of the scalar field $\hat \phi$ (see Fig. \ref{fig:phi_min}) to its zero temperature VEV~(\ref{eq:phihat0}). This dynamically generates and additional source of misalignment, $\hat \phi_V$, which we refer to as VEV misalignment.  
\end{itemize}

The final misalignment is in general influenced by all of the above contributions. 
Note that both thermal misalignment and VEV misalignment are dynamical processes occurring during the radiation era. 
As we will see, these sources of misalignment depend on the DM model parameters, i.e., the mass $m_\phi$ (or equivalently $\kappa$) and the coupling $A$ (or equivalently $\beta$). On the other hand, the initial value $\hat \phi_{i}$ after reheating depends on the detailed dynamics 
during inflation and can in principle be arbitrary.
An important question then is to what extent the relic abundance prediction depends on the assumed initial condition. If the relic density prediction is insensitive to the initial conditions and only depends on the DM model parameters, this opens up the exciting prospect of confronting the cosmological production mechanism with experiments, since the signatures and predictions for the latter are determined by the same model parameters (mass and coupling). 

As we will show in detail in Sec.~\ref{sec:Reg1}, for larger scalar masses, $m_\phi \gtrsim$ few meV ($\kappa \gtrsim 10^{3}$), thermal misalignment dominates over VEV misalignment, $\hat \phi_{T} \gg \hat \phi_{V}$. Then provided $\hat \phi_{i} \ll  \hat \phi_{T}$, which can be satisfied for a broad range of conceivable initial field values given Eq.~(\ref{eq:phiT}), the thermal misalignment generated during the radiation era will overwhelm (or ``erase'') the contribution from the initial field value. 
If this condition holds, the late-time oscillation amplitude and resulting scalar relic abundance is controlled entirely by the DM model parameters. We can then define a relic density target corresponding to a line in the $m_\phi-A$ plane yielding a DM abundance in accord with the measured value, $\Omega_{\rm DM} \simeq 0.26$~\cite{Planck:2018vyg}. 
Modulo fine tuning of initial conditions, parameters above the relic density target line will lead to even larger scalar oscillation amplitudes generated by thermal misalignment and would thus overclose the universe. 
These features, such as correlations between DM relic density and its mass and coupling, as well as insensitivity to initial conditions and UV physics, are reminiscent of similar attractive aspects found in models of weakly-interacting-massive particle (WIMP) DM. 

For lower masses, $m_\phi \lesssim  10^{-5}$ eV ($\kappa \lesssim 1$), VEV misalignment dominates over thermal misalignment, $\phi_{V} \gg \phi_{T}$. As will be clearly shown in Sec.~\ref{sec:relic-estimate}, the insensitivity of the scalar relic density to the initial field value $\hat \phi_{i}$ does not carry over to this lower mass regime.

Below we will detail two choices of initial conditions that will be used in our numerical and analytic estimates in the coming sections. 
These two choices will allow us to demonstrate the insensitivity of the relic density to initial conditions for scalars with relatively large masses where thermal misalignment dominates, as well as study the impact of different initial conditions for scalars with relatively low masses where VEV misalignment is important.

\begin{itemize}
    \item $\phi_{i} = \phi_{0}$:

    In this case, the scalar is initially at  its zero-temperature VEV, $\phi_0$, given by Eq.~(\ref{eq:vphi0}). 
    This initial condition is naturally realized through a
    sufficiently long enough period of inflation with a low enough inflationary Hubble parameter, $H_{I} \ll v$. In this case,
    electroweak symmetry is broken during inflation, the Higgs is close to its VEV $v$, and the $\phi$ VEV is approximately at $\phi_{0}$.
    Given a long enough period of inflation the distribution of $\phi$ will relax to the average $\langle \phi \rangle|_{H_{I}} = \phi_{i} = \phi_{0}$. 
    This relaxation will take order $\mathcal{N} \sim H^{2}_{I}/m^{2}_{\phi}$ $e$-folds, and the variance of $\phi$ will be given by $\sigma_{\phi_{i}} = 3 H^{4}_{I}/(8\pi m^{2}_{\phi})$.
    Note that $H_{I}$ cannot be arbitrarily low if thermal misalignment is to be important, which requires the reheat temperature to satisfy $T_{\rm RH} \gg v$, though this condition can easily be met.
    We also require $\sigma_{\phi_{i}} \ll \phi_{T}$, $(\phi_{V})$ for the thermal (VEV) misalignment to be important. 
    We note that a low inflationary Hubble scale naturally suppresses scalar fluctuations (see, e.g., Refs.~\cite{Graham:2018jyp,Takahashi:2018tdu,Tenkanen:2019aij}), thus easing otherwise stringent CMB constraints on isocurvature  perturbations~\cite{Planck:2018jri}. 

    As emphasized above, for heavier scalars the displacement generated by thermal misalignment overwhelms the initial field value assumed here. 
    On the other hand, for lighter scalars thermal misalignment is negligible and VEV misalignment operates, though we must carefully consider the impact of the assumed initial condition. The scalar is held fixed by Hubble friction for some time near its initial value, $\hat{\phi_{i}} \sim \hat{\phi_{0}}$. 
    Then at temperatures just above the EWPT, the $\hat \phi$ VEV adjusts to field values that are much smaller than its initial value
    $\hat \phi_0$, as can be seen in in Fig.~\ref{fig:phi_min}. 
    At this stage, the scalar field slow-rolls away from its initial position, generating misalignment. After the EWPT, the location of the $\hat \phi$ minimum quickly transitions to its zero-temperature VEV $\hat \phi_0$, and eventually oscillations commence after the expansion rate drops below the scalar mass. It will be shown that the parametric dependence of the VEV misalignment for this inital condition is 
     \begin{align}
            \hat \phi_{V} \propto \beta,  ~~~~~~~~  (\hat \phi_i = \hat \phi_0) \,.
        \end{align}
    A more thorough treatment of this dynamics will be presented in Sec.~\ref{sec:Reg3}.

 \item $\phi_{i} = 0$:

    This initial condition is chosen as a representative example of the general situation where 
     $|\phi_{i}|$ is vastly different than $|\phi_{0}|$.  
     For larger scalar masses, thermal misalignment again dominates the evolution and the scalar relic abundance is insensitive to the assumed initial field value.
     In contrast, for smaller scalar masses, 
     VEV misalignment dominates over thermal misalignment, although we must again understand the role of the initial condition.
     In fact, the evolution of the scalar field in this case is quite simple. Initially, the scalar is held at its initial location at the origin by Hubble friction. The EWPT triggers a shift in the $\phi$ VEV to its zero-temperature value $\phi_0$, generating VEV misalignment of parametric size (see Sec.~\ref{sec:Reg3}):
       \begin{align}
    \hat{\phi}_{V} 
    \simeq -\frac{\beta^{2}}{2\kappa^{2} \lambda}~~~~~~~~  (\hat \phi_i = 0) \,.
    \end{align}
     At some later time the Hubble rate falls below the scalar mass and oscillations begin. 

\end{itemize}

It can be seen in Fig.~\ref{fig:relic} that the relic density lines  in the lower scalar  mass region are strongly sensitive to the initial field value, while instead those in the higher scalar mass range are the same for both assumed initial conditions. 

\subsection{Scalar field dynamics}
\label{sec:scalar-dynamics}
We are now ready to discuss the evolution of the scalar field $\phi$, which is governed by the following equation of motion during the radiation-dominated era
\begin{align}
\Ddot\phi+3H\dot{\phi}+\frac{\partial V_{\rm eff}}{\partial\phi}=0\,,
\label{eq:phi-EOM}
\end{align}
where again $V_{\rm eff} = \mu^{4} \, \hat{V}_{\rm eff}$ is given in Eqs.~(\ref{eq:Veff},\ref{eq:Vhat}) and the Hubble parameter $H$ has been defined at the beginning of Sec.~\ref{sec:Veff}. 

For simplicity, we will ignore the small time variation of $g_{*}$ during the initial phase of the $\phi$ evolution until $\phi$ oscillations begin, fixing it to the high temperature SM value of $g_*^{\rm SM} = 106.75$ ($\gamma =\sqrt{\pi^2 g_*/90} \simeq 3.4$). 
For large scalar masses, oscillations begin well before the EWPT, and as such $g_* =g_*^{\rm SM}$ is exactly correct. 
Instead, for smaller masses, oscillations begin below the EWPT
and $g_{*}$ starts to decreases as $t,W,Z,h$ fall out of thermal equilibrium, giving for example $g_{*} \sim 80$ at $y \sim 0.01$.  
Thus for smaller masses our constant $g_{*}= g_*^{\rm SM}$ approximation will result in a $\mathcal{O}(10\%)$ error in the relic abundance prediction.

Let us also briefly discuss a potential additional source of scalar field damping. As the scalar field evolves, the effective mass of the Higgs particles in the bath changes though their momenta does not, causing a small deviation in the Higgs phase space distribution from its equilibrium value. This can manifest as an additional effective friction term in the scalar equation of motion; see e.g., Ref.~\cite{Yokoyama:1998ju}. However, the Higgs quickly relaxes toward equilibrium through fast number changing processes, such that this source of damping is much smaller than the other terms in Eq.~(\ref{eq:phi-EOM}). See also Ref.~\cite{Bodeker:2022ihg} for discussion.

We can re-express the $\phi$ equation of motion (\ref{eq:phi-EOM}) in terms of the dimensionless variables defined in Eq.~(\ref{eq:dimless}), with independent variable of time $t$ traded for $y = T/\mu$. We obtain 
\begin{align}
\label{eq:mastereq}
    \hat{\phi}''+\frac{1}{\gamma^{2}y^{6}}\left[\kappa^{2}\hat{\phi}+\frac{\beta \hat{h}^{2}}{2}+\frac{\beta y^{2}}{2\pi^{2}}\left(J'_{B}[\eta_{h}]+3J'_{B}[\eta_{\chi}]\right)\right]=0\,.
\end{align}
Here, $\hat{h}=\hat h(\hat \phi, y)$ is the solution to Eq.~(\ref{eq:hath}), as discussed in detail in Sec.~\ref{sec:EWPT}. We then solve the $\hat \phi$ equation of motion (\ref{eq:mastereq}) subject to the initial conditions as described in the previous section. These are defined at $y_i = T_i/\mu$, where $y_{i} \gg 1$ corresponds to some suitably early time well before the scalar starts to oscillate.
The scalar field is evolved until $y_f = T_f/\mu$ corresponding to some suitable time after it begins oscillating and redshifts as matter. 
From this time on, the $\phi$ comoving energy density, $\rho_\phi/s$, where 
$\rho_\phi$ is the $\phi$ energy density and $s(T)=(2\pi^2/45) g_{*S}(T) T^3$ is the total entropy density of the plasma, is conserved. %
Using this fact, we then arrive at our estimate for the $\phi$ energy density in the present epoch,
\begin{align}
\label{eq:phi-relic-density}
    \rho_{\phi,0} = \frac{ g_{*S}(y_0) }{g_{*S}(y_f)} \left(\frac{y_0}{y_f}\right)^3 \rho_\phi(y_f)\,,
\end{align}
where $y_0 = T_0/\mu$ with $T_0 \simeq 2.7$ K. We use $g_{*S}(y_f) = 106.75$ and $g_{*S}(y_0) \simeq 3.91$ in our estimate. 
Expressing the relic density in terms of the density parameter $\Omega_{\phi,0} = \rho_{\phi,0}/\rho_{c,0}$, with $\rho_{c,0} = 3 M_{\rm pl}^2 H_0^2$ the critical density today, we show the parameters leading to the observed DM relic abundance, $\Omega_{\phi,0} = \Omega_{\rm DM} \simeq 0.26$~\cite{Planck:2018vyg} in Fig.~\ref{fig:relic}.

Before discussing these results and comparing with the various experimental probes of the model, it is valuable to gain analytical insight into the $\phi$ evolution and resulting relic density estimate, which we explore in the next section. 

\subsection{Relic Density Estimation} 
\label{sec:relic-estimate}

As the universe expands, the Hubble parameter decreases until it eventually falls below the effective $\phi$ mass, marking the onset of scalar oscillations. This can be quantified by the condition $[3H(y_{\rm osc})]^2 = m_{\phi}^{2}(y_{\rm osc}) \simeq m_\phi^2$, which gives the oscillation temperature as 
\begin{align}
\label{eq:yosc}
y_{\rm osc} = \frac{T_{\rm osc}}{\mu} = \sqrt{\frac{\kappa}{3\gamma}} \, . 
\end{align}
Here we have used the fact that the effective squared scalar mass is dominated by the tree-level contribution $m_\phi^2$ over the entire parameter space we consider and for all temperatures. 
Several epochs after the oscillations begin, we can safely calculate the energy density stored in the scalar field and then redshift it to present times, which will yield the $\phi$ relic density given by Eq.~(\ref{eq:phi-relic-density}). 
Typically, once the scalar starts to oscillate, it behaves like DM with an oscillation amplitude set by
\begin{align}
    \phi_{\rm osc} \equiv \phi(y_{\rm osc}) \, , 
\end{align}
i.e., the scalar amplitude at the beginning of oscillations. 
To determine $\phi_{\rm osc}$ we must study the evolution of the scalar field during the radiation era up until the onset of oscillations. As discussed at length in Sec.~\ref{sec:ic}, there are additional dynamical sources of misalignment present in the Higgs portal model and contribute to the oscillation amplitude. For large scalar masses, thermal misalignment provides the main contribution, while for small scalar masses VEV misalignment (along with initial conditions) is dominant. In the intermediate mass range, both effects are important. 
In the following, we will therefore divide  the parameter space into distinct regions characterized by which sources of misalignment dominate and then develop a suitable analytic estimate for the $\phi$ evolution and relic abundance. 
These estimates will be compared with our relic density prediction obtained from the exact numerical solution of the $\phi$ equation of motion~(\ref{eq:phi-EOM}), as described in the previous section.

\subsubsection{\textbf{Region I $(\kappa \gtrsim 10^{3},~ m_{\phi} \gtrsim 3 \times 10^{-3}\, {\rm eV})$}}
\label{sec:Reg1}

Region I is defined to be the region of large scalar masses $m_\phi$ (large $\kappa$). In this region, the the amplitude of oscillations at $y_{c}$ due to thermal misalignment dominates over the ``kick'' imparted to the scalar 
resulting from the EWPT (VEV misalignmnet).
To understand this point, we first need to estimate the thermal misalignment generated at high temperatures by considering Eq.~(\ref{eq:mastereq}). For $y \gg y_{\rm osc} \gg y_{c}$, $\hat{h}(\hat \phi,y) = 0$ and 
we can ignore the $\kappa^{2} \hat{\phi}$ term since it is subdominant to the finite-temperature contribution to the equation of motion. Hence, in this regime, Eq.~(\ref{eq:mastereq}) takes the approximate form
\begin{align}
    \label{eq:reg1mastereq}
    \hat{\phi}''(y) + \frac{\beta}{2\pi^{2}\gamma^{2}y^{4}}\left(J'_{B}[\eta_{h}]+ 3 (J'_{B}[\eta_{\chi}] \right) &= 0, \nonumber \\
   \Longrightarrow~~~~ \hat{\phi}''(y) + \frac{\beta}{\pi^{2}\gamma^{2}y^{4}} &= 0\,,
\end{align}
where in the second line we have used the fact that $ J'_{B}[\eta_{h}]+ 3 J'_{B}[\eta_{\chi}] \approx 2$ for $y\gg y_{c}$ (see App.~\ref{sec:thermal-masses} for definitions of $\eta_{h},\eta_{\chi}$). 
The solution to Eq.~(\ref{eq:reg1mastereq}), assuming negligible initial field velocity, is 
\begin{align}
    \label{eq:reg1evol}
    \hat{\phi}(y) = -\frac{\beta}{6 \pi^{2} \gamma^{2} y^{2}} + \hat{\phi}_{i}\,,
\end{align}
where the first term represents the thermal misalignment, $\hat \phi_{T}(y) = -\beta/(6 \pi^{2} \gamma^{2} y^{2})$, while the second is the initial condition $\hat{\phi_{i}}$. As the temperature drops below $y_{\rm osc} = \sqrt{\kappa/3\gamma}$, the mass term $(\kappa^{2} \hat{\phi})$ in Eq.~(\ref{eq:mastereq}) begins to dominate over the thermal contribution and the scalar starts to oscillate and redshift like DM. 
Plugging  $y = y_{\rm osc}$ into Eq.~(\ref{eq:reg1evol}), we find the scalar amplitude at the onset of oscillations is given by
\begin{align}
    \label{eq:reg1sol}
   \hat \phi_{\rm osc} =  \hat{\phi}(y_{\rm osc}) = -\frac{\beta}{2 \pi^{2} \gamma \kappa} + \hat {\phi}_{i} \, .
\end{align}
For the first initial condition $ \hat{\phi}_{i} =  \hat{\phi}_{0} \simeq -\beta/(2\lambda\kappa^{2}) $ motivated in Sec.~\ref{sec:ic}, we observe that thermal misalignment $\hat{\phi}_{T}$ dominates over $\hat{\phi}_{i} = \hat{\phi}_{0}$ for $\kappa \gtrsim  \pi^{2}\gamma/\lambda \approx 300$.
Thus, our definition of Region I, $\kappa > 10^{3}$, satisfies this criterion. 

\begin{figure}
    \centering
        \includegraphics[width=0.495\columnwidth]{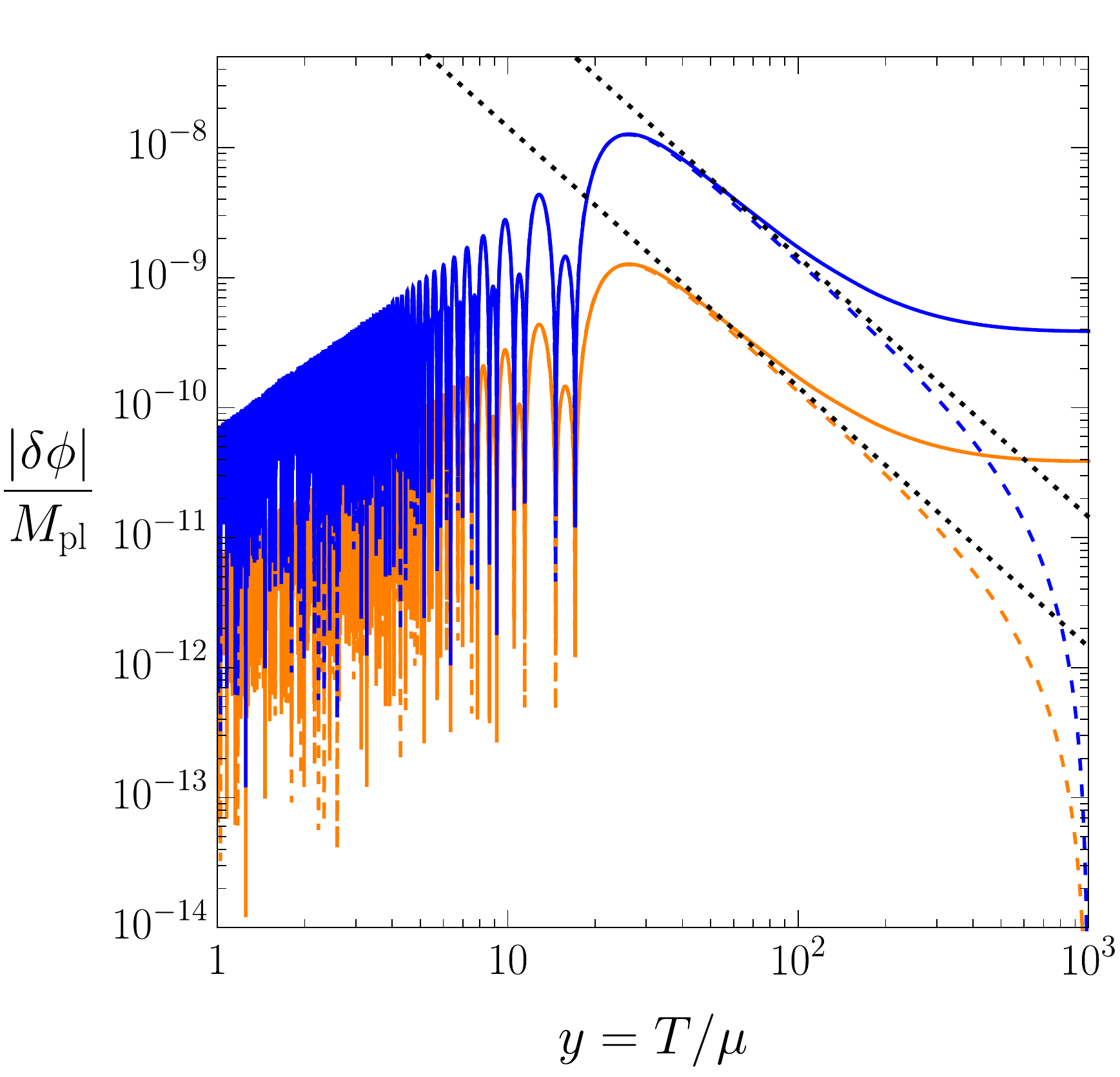}
\caption{Scalar field evolution $\hat \phi(y)$ relative to its temperature-dependent Higgs-induced minimum, $\delta\phi(y) \equiv \hat \phi(y) - [ -\beta \hat h_0^2(y)/(2 \kappa^2)]$  
 for two benchmark models in Region I: 
   $\beta=10^{-2}, \kappa=10^4$ (blue) and $\beta=10^{-3}, \kappa=10^4$ (orange). The black dotted lines show the corresponding approximate initial thermal misalignment trajectories of Eq.~(\ref{eq:reg1sol}). The $\beta^2$ scaling due to thermal misalignment is clearly observed. 
   The evolution is shown for the two choices of initial conditions discussed in Sec.~\ref{sec:ic}: $\hat \phi_i= \hat \phi_0$ (solid) and $\hat{\phi}=0$ (dashed).
    It is evident that the late time $\hat \phi$ evolution is independent of the initial value of the scalar field for these choices. 
    }
    \label{fig:Reg1_phi}
\end{figure}

In Fig.~\ref{fig:Reg1_phi} we display the numerical evolution of $\hat \phi(y)$ 
relative to its temperature-dependent Higgs-induced minimum, 
$\delta\phi(y) \equiv \hat \phi(y) - [ -\beta \hat h_0^2(y)/(2 \kappa^2)]$,  
for two benchmark models in Region I and the initial conditions $ \hat{\phi}_{i} = \hat{\phi}_{0}$ and $\hat{\phi}_{i} = 0$ discussed in Sec.~\ref{sec:ic}. 
This quantity provides a convenient visualization of both the thermal misalignment generated at high temperatures as well as the negligible impact of VEV misalignment for $y < y_c$. 
The temperature-dependent Higgs-induced minimum, $[ -\beta \hat h_0^2(y)/(2 \kappa^2)]$, is simply the first term in the full temperature-dependent minimum $\hat \phi_{\rm min}$ in Eq.~(\ref{eq:phimin}), which dominates the expression at low temperatures, i.e.,
\begin{align}
\label{eq:phi-min-y}
\hat \phi_{\rm min}(y) 
\simeq - \frac{\beta \hat h_0^2(y)}{2 \kappa^2}\, ~~~~~~~~(y \lesssim y_c).
\end{align}
This is also clear from Fig.~\ref{fig:phi_min}.
The dominance of thermal misalignment and the independence of the the oscillation amplitude on the initial conditions are clearly seen in Fig.~\ref{fig:Reg1_phi}. 

We will now justify the condition $\kappa \gtrsim 10^{3}$ used to define Region I.
For $\kappa \gtrsim 10^{3}$, oscillations of the scalar field around its high temperature minimum begin at $y_{\rm osc} \gg y_{c}$. 
The amplitude of the oscillations then decreases with temperature as $y^{3/2}$ (i.e., $\phi$ redshifts like matter) until $y = y_{c}$. At $y_{c}$ the Higgs field transitions nearly instantaneously, causing a shift in the minimum of $\hat{\phi}$ from the its minimum at $y>y_{c}$
to $\hat{\phi}_{0}$. 
If the change in the $\hat{\phi}$ minimum near $y = y_c$ 
within one oscillation time period is larger than the oscillation amplitude, then the relic density is controlled by 
VEV misalignment. 
Conversely, if the oscillation amplitude at $y = y_c$ is much larger than
the shift in the $\hat{\phi}$ minimum due to the EWPT, then thermal misalignment dominates. 
For large $\kappa$ (large $m_{\phi}$), the latter scenario of thermal misalignment applies. In what follows, we will derive a lower bound on $\kappa$ above which the thermal misalignment dominates (defined to be Region I) and then develop an analytical approximation for the $\phi$ relic density.

At $y_{c}$, the minimum of $\hat{\phi}$ transitions from being governed by the thermal contribution to the potential
to being controlled by the Higgs VEV $\hat{h}$ (see Fig. \ref{fig:phi_min}). 
In this region, the $\hat \phi$ minimum swiftly changes according $\hat {\phi}_{\rm min}$ given in Eq.~(\ref{eq:phi-min-y}). 
This rapid shift in $ \hat{\phi}_{\rm min}$ provides a ``kick'' by effectively changing the amplitude of the oscillation. 
The magnitude of this kick is the variation of $\hat{\phi}_{\rm min}$ within one half oscillation of the scalar field i.e. within $\Delta t = \pi/m_{\phi}$. It can be shown that this time period corresponds to $\Delta y = - y_c^3 \gamma \pi/\kappa$.
The change in the $\hat \phi$ minimum is then given by
\begin{align}
    \label{eq:kick}
    \Delta \hat{\phi}_{\rm min}  &\simeq \frac{\partial\hat{\phi}_{\rm min}}{\partial y} \Delta y  = - \frac{\beta}{2\kappa^{2}}\frac{\partial \hat{h}^{2}}{\partial y} \Delta y \nonumber \\
    & \simeq -\frac{\pi \gamma y_c^4 \beta }{2 \lambda \kappa^{3} },
\end{align}
where in the last line we have used $\partial \hat{h}^{2}/\partial y \simeq -y_{c}/\lambda$ (this can be seen from the evolution equation for $\hat{h}^{2}$ in Eq.~(\ref{eq:hath})).
This is to be compared to the oscillation amplitude near $y_c$ due to thermal misalignment, which is given by
\begin{align}
    \label{eq:ycamp}
    \hat{\phi}_{\rm osc}(y_{c}) = \hat{\phi}(y_{\rm osc}) \left(\frac{y_{c}}{y_{\rm osc}}\right)^{3/2} = -\frac{(\sqrt 3 y_{c})^{3/2}\beta }{2\pi^{2} \gamma ^{1/4} \kappa^{7/4}}\,,
\end{align}
where we have used Eq.~(\ref{eq:reg1sol}) and redshifted it to $y_{c}$. For Region I, we require thermal misalignment to dominate over the ``kick'', taking as our criterion $\hat{\phi}_{\rm osc}(y_{c}) > 3 \Delta \hat{\phi}_{\rm min}$.  Using Eqs.~(\ref{eq:kick},\ref{eq:ycamp}) we find that this is satisfied for
\begin{align}
    \label{eq:reg1boundary}
    \kappa > \gamma y_c^2 \left(\frac{3 \pi^{12}}{\lambda^4 }\right)^{1/5}  \simeq 10^3\,.
\end{align}
In practice we find that numerically we agree well with the estimates of Region I for $\kappa > 10^{3}$. We thus take $\kappa = 10^{3}$ as the boundary for Region I.

Having established that thermal misalignment dominates in Region I, $\kappa \geq 10^3$, we are now ready to provide an analytic estimate of the $\phi$ relic abundance. We use Eq.~(\ref{eq:phi-relic-density}) to estimate $\rho_{\phi,0}$, the $\phi$ energy density today. 
As input to this equation, we take $y_f = y_{\rm osc}$ as the temperature at which oscillations begins, with  $y_{\rm osc}$ given by Eq.~(\ref{eq:yosc}). We also use $\rho(y_f) = \rho(y_{\rm osc}) \simeq \tfrac{1}{2} m_\phi^2 \phi(y_{\rm osc})^2$, with  $\phi(y_{\rm osc})$ given by the first term in Eq.~(\ref{eq:reg1sol}). Other inputs needed are described below Eq.~(\ref{eq:phi-relic-density}). Putting everything together, we arrive at the following estimate for the DM density parameter,
\begin{align}
    \label{eq:ODMreg1}
    \Omega_{\phi,0}
    &\simeq  \frac{g_{*S}^0}{g_{*S}^{\rm osc}}\, \frac{y_0^3 \, \mu^4}{ \rho_{c,0}} \,
    \frac{3^{3/2}}{8 \pi^4} \, \frac{\beta^2 }{ \gamma^{1/2} \kappa^{3/2}} \nonumber \\
    &\simeq 0.26 \left(\frac{\beta}{0.05}\right)^{2} \left(\frac{1000}{\kappa}\right)^{3/2}\,.
\end{align}
Parameters leading to the observed DM abundance according to the approximate analytic estimate, Eq.~(\ref{eq:ODMreg1}), are displayed as a dashed line in the $m_\phi - A$ plane in Fig.~\ref{fig:relic} for $m_\phi \gtrsim 10^{-3}$ eV, agreeing well with the calculation using the exact numerical solution to the $\phi$ equation of motion~(\ref{eq:phi-EOM}) in Region I. 
Eq.~(\ref{eq:ODMreg1}) (and the associated numerical estimate) represents one of the most important results of this work, providing a prediction for the observed DM relic density in terms of only the DM model parameters ($m_\phi$ and $A$)
which is insensitive to initial conditions provided $|\phi_i| \ll |\phi_T|$, 
as discussed in Sec.~\ref{sec:ic}. 
In particular, note that for parameters above the relic density line in Fig.~\ref{fig:relic}, initial conditions with $|\phi_{i}| \gg |\phi_{T}|$ will overclose the universe. 
Since thermal misalignment is unavoidable given a standard radiation dominated cosmology and a high enough reheating temperature $T_{RH} \gg v$,
this gives a cosmologically motivated target in the parameter space for experiments to pursue. In fact, requiring DM not to overclose the universe currently provides the best constraint on the model (with the caveats mentioned above) over the vast majority of the parameter space. 

\subsubsection{\textbf{Region II  $(10 \lesssim \kappa \lesssim 10^{3},~  3\times 10^{-5} \lesssim m_{\phi} \lesssim 3 \times 10^{-3} {\rm eV})$}} 
\label{sec:Reg2}

We now move to Region II, corresponding to the intermediate mass range 
$(10 \lesssim \kappa \lesssim 10^{3},~  3\times 10^{-5} \lesssim m_{\phi} \lesssim 3 \times 10^{-3} {\rm eV})$.
In this region, the final misalignment amplitude is influenced in an important way by the change in the minimum of $\phi$ triggered by the EWPT at $y = y_c$. 

\begin{figure}
    \centering
    \includegraphics[width=0.493\columnwidth]{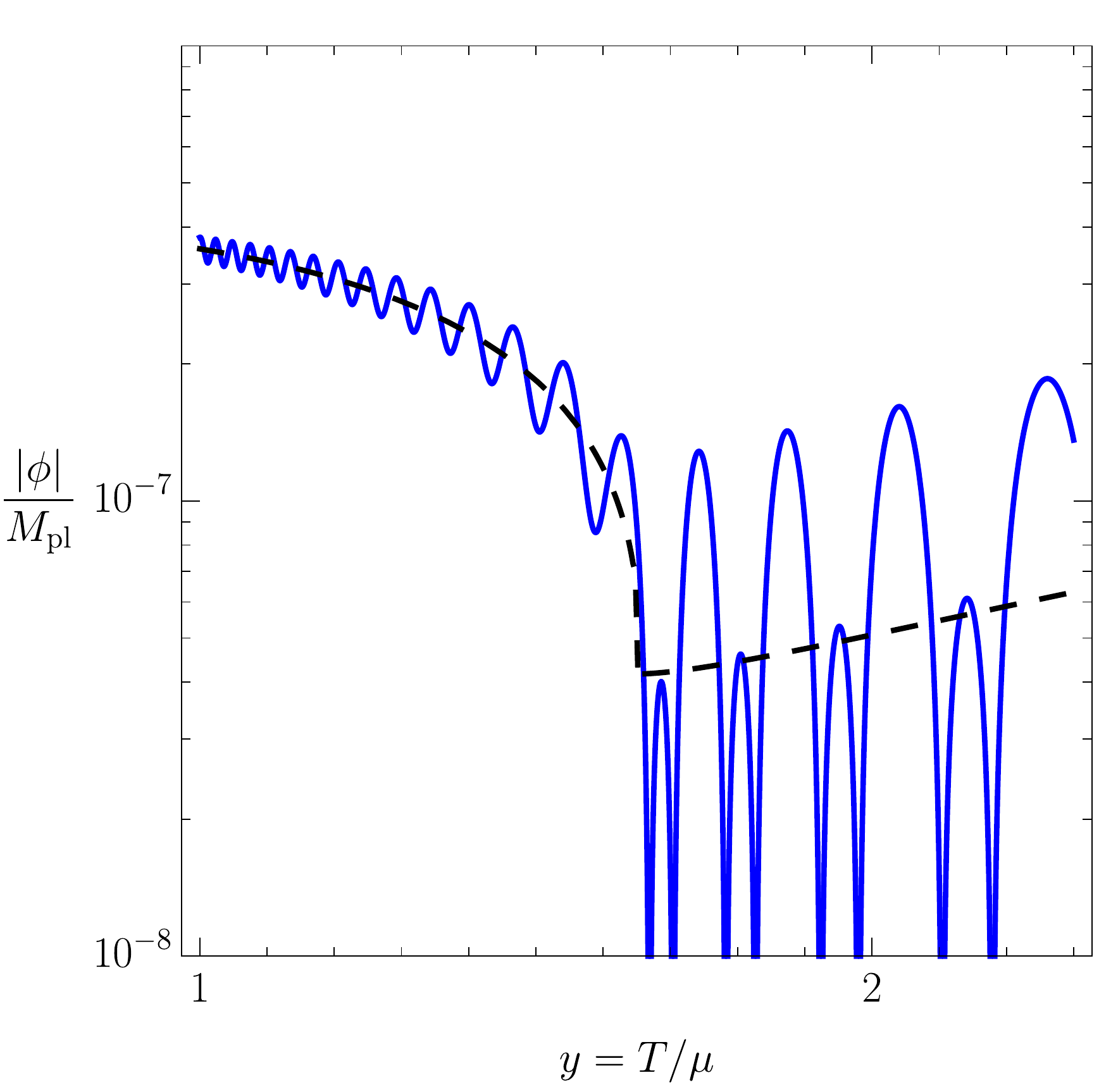}~~~~~
    \includegraphics[width=0.493\columnwidth]{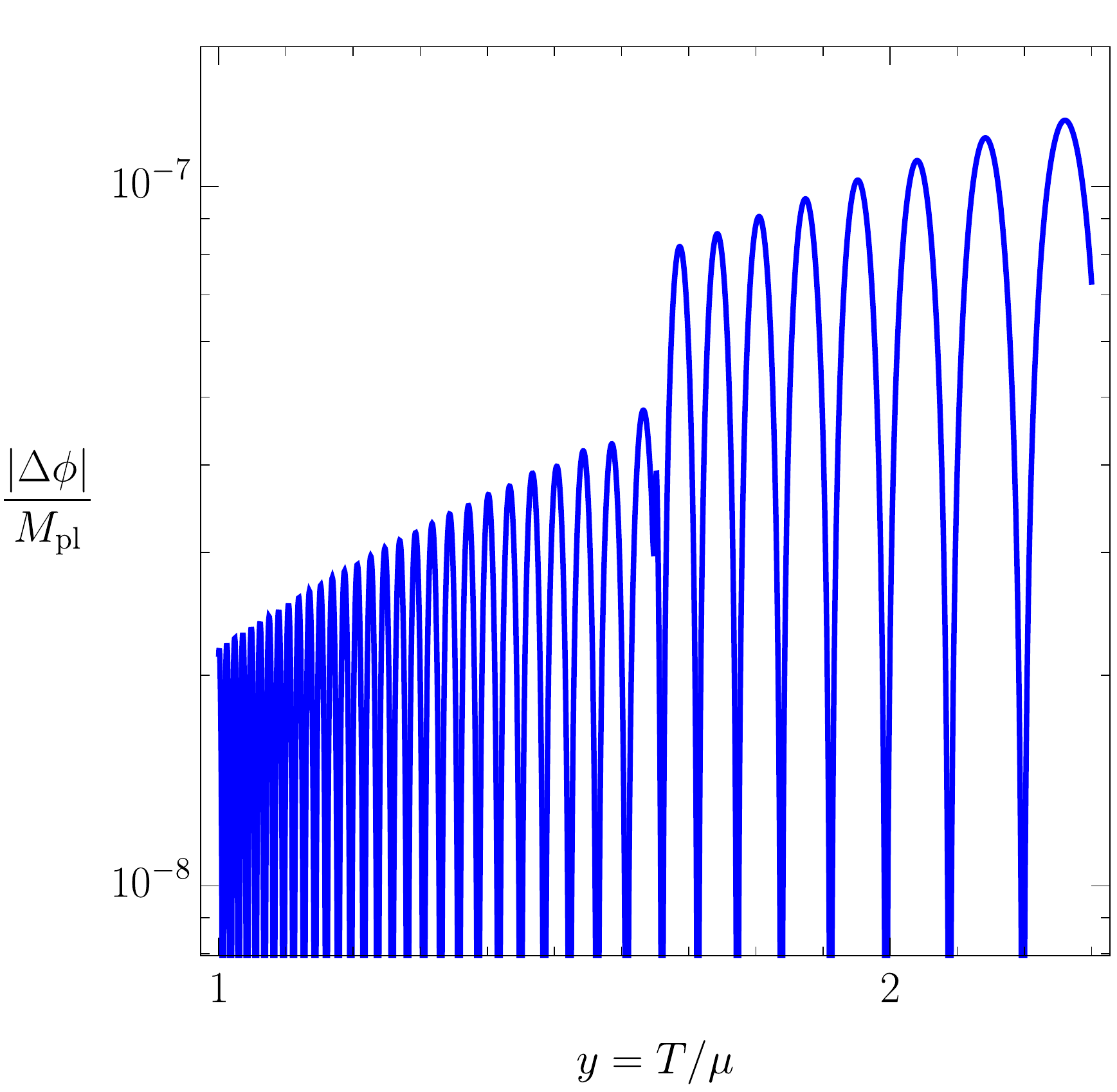}
    \caption{\textbf{Left panel:} The evolution of $\phi$ (blue) along with the progression of the temperature-dependent minimum of $\phi$ (black dashed) during the EWPT. The true minimum jumps quickly after the transition, reducing the effective oscillation amplitude.
    \textbf{Right panel:} This amplitude reduction is clearly seen in $\Delta \phi(y) \equiv \phi(y) - \phi_{\rm min}(y)$, which shows the oscillations relative to the temperature dependent $\phi$ minimum, Eq.~(\ref{eq:phi-min-y}).
    In both plots we have chosen the benchmark parameters $\kappa = 10^{2.985}$, $\beta = 10^{-1}$.
    }
    \label{fig:peak_phi}
\end{figure}

Notably, for masses near the upper end of Region II, $m_\phi \sim {\cal O}(10^{-3} {\rm eV})$, a competition between thermal misalignment and VEV misalignment leads to a novel forced resonance phenomena which impacts the relic abundance in a rather dramatic fashion. At the EWPT, the Higgs field rapidly moves from the origin towards $h = v$, 
simultaneously inducing a shift the $\phi$ VEV towards its zero-temperature value $\phi_0$. 
The acts as a step-like forcing term in the $\phi$ equation of motion (\ref{eq:mastereq}), causing a suppression or enhancement in the oscillation amplitude depending on the relative phase between the oscillations and Higgs source term, which in turn depends on the the scalar mass. 
For instance, if the scalar is near its oscillation maximum as this shift happens, the effective oscillation amplitude is reduced, thus requiring a larger coupling $\beta$ to produce the observed DM abundance.
This behavior is shown in Fig. \ref{fig:peak_phi} for one example benchmark point.
A striking series of recurring peaks and valleys in the coupling $\beta$ yielding the correct DM abundance is observed as the scalar mass is varied, as shown the inset of Fig.~\ref{fig:relic}.

A potential caveat to these results is that they rely on the precise description of the Higgs field evolution through the EWPT. While we have modeled the Higgs evolution using the effective potential, Eq.~(\ref{eq:Vhat}), a more accurate description would incorporate the results from lattice studies of the EWPT~\cite{Kajantie:1996mn,DOnofrio:2015gop}, which is beyond our present scope. This could influence the detailed nature of the scalar field evolution and forced resonance phenomena outlined above.

For masses near the lower end of Region II, VEV misalignment dominates and the predicted DM abundance starts to be sensitive to initial conditions, which is also evident from Fig.~\ref{fig:relic}. The impact of VEV misalignment and the role of the initial conditions for the low mass scalars will be clarified in the next through our analysis of Region III. 

\subsubsection{\textbf{Region III $(\kappa \lesssim 1, m_{\phi} \lesssim  3 \times 10^{-6} \, {\rm eV})$} }
\label{sec:Reg3}

Finally, we come to Region III, which may roughly be defined by the condition that oscillations begin below the EWPT, $y_{\rm osc} \lesssim 1 \implies \kappa \lesssim 3\gamma$. We will simply take $\kappa \lesssim 1$ ($m_{\phi} \lesssim  3 \times 10^{-6}\, {\rm eV}$) for concreteness. 
In this low mass regime, VEV misalignment caused by the EWPT dominates over thermal misalignment. However, the final $\phi$ misalignment and number density depends sensitively on the initial conditions (see Fig. \ref{fig:relic}). In the following, we will examine two distinct choices for the initial condition,  $\hat {\phi}_i = \hat {\phi}_0$ and $\hat \phi_i = 0$, as discussed in Sec.~\ref{sec:ic}.

\begin{figure}
    \centering
    \includegraphics[width=0.5\columnwidth]{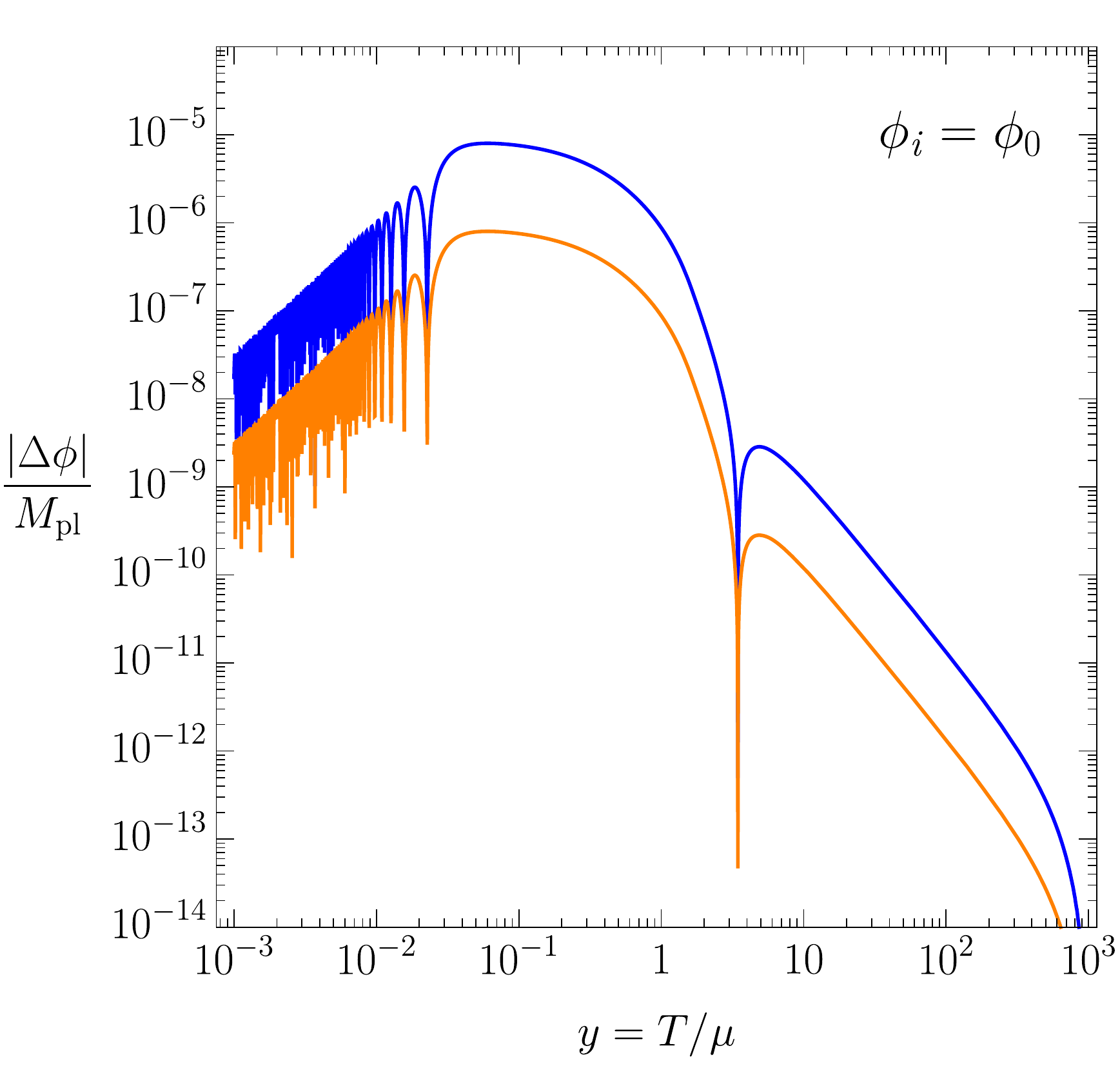}~~~~~~
    \includegraphics[width=0.47\columnwidth]{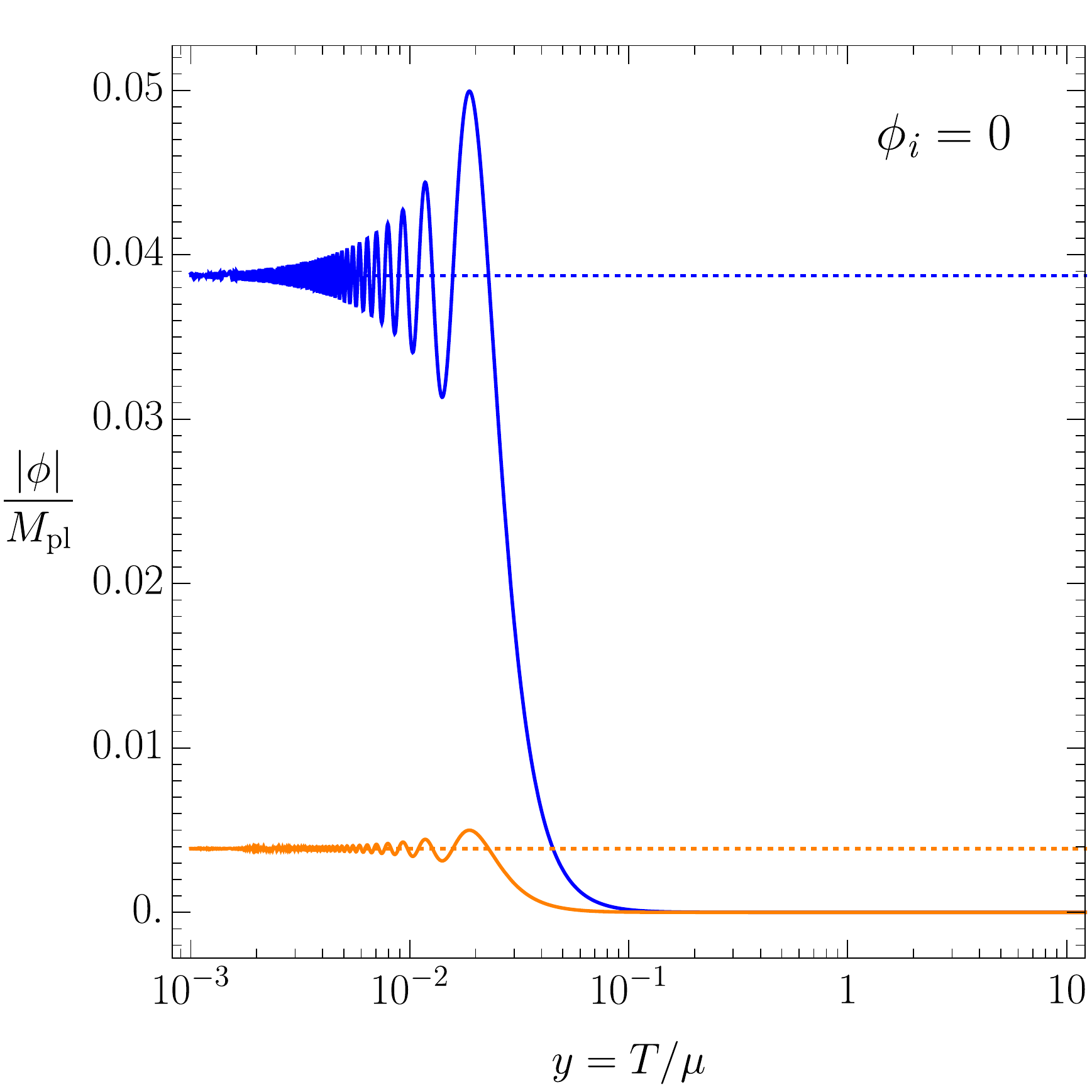}
    \caption{Scalar field evolution in Region III, where VEV misalignment is important, for two choices of initial conditions. 
    \textbf{Left panel:} The initial condition is chosen to be $\phi_i = \phi_0$. 
    The evolution $\Delta \phi(y) \equiv \phi(y) - \phi_0$ is shown for two benchmark models: $\kappa=10^{-2}$, $\beta=10^{-4}$ (blue curves), and $\kappa=10^{-2}$, $\beta=10^{-5}$ (orange curves).
    \textbf{Right panel:} The initial condition is chosen to be $\phi_i = 0$.
    The evolution $\phi(y)$ is shown for two benchmark models: $\kappa=10^{-2}$, $\beta=10^{-6}$ (blue curves), and $\kappa=10^{-2}$, $\beta=10^{-7}$ (orange curves). The dotted lines show the corresponding zero temperature minima $\phi_0$.
    } 
    \label{fig:Reg2_phi}
\end{figure}

{\bf (i)  $\hat {\phi}_{i} = \hat {\phi}_{0}$}: ~~
Let us first consider the initial condition $\hat {\phi}_{i} = \hat {\phi}_{0} \simeq -\beta/(2 \lambda \kappa^{2})$.
Examining the $\hat{\phi}$ equation of motion (\ref{eq:mastereq}) for $\hat{\phi} \approx \hat{\phi}_{0}$, we observe that the mass term $\kappa^{2} \hat{\phi}$ dominates over the thermal term for $\frac{y^{2}}{2\pi^{2}} \beta \left(J'_{B}[\eta_{h}]+ 3 J'_{B} [\eta_{\chi}]\right)$ for $y^{2} \lesssim \pi^{2}/(2\lambda)$. Thus, the scalar field first experiences thermal misalignment 
$\hat{\phi}_{T}  \sim - \lambda \beta/(3\pi^{4} \gamma^{2})$, 
where we have used Eq.~(\ref{eq:reg1sol}) and $y^{2} \sim \pi^{2}/(2\lambda)$. 
However, we note that the displacement of the scalar from $\hat \phi_0$ generated by thermal misalignment is minuscule, $|\hat{\phi}_{T}| \ll \hat{\phi}_{0}$, 
since $\kappa \lesssim 1$ in this region.
For $y_{c} <y \lesssim \sqrt{\pi^{2}/(2\lambda)} \simeq 7$ 
the mass term dominates and the equation of motion becomes
\begin{align}
    \label{eq:reg2eom}
    \hat{\phi}''(y) + \frac{1}{\gamma^{2} y^{6}}\left(\kappa^{2} \hat{\phi}\right) = 0\,, \nonumber \\
 \Longrightarrow~~~~    \hat{\phi}''(y) - \frac{\beta}{2\gamma^{2} \lambda} \frac{1}{y^6} = 0\, ,
\end{align}
where in the second line we have assumed $\hat{\phi} \simeq \hat{\phi}_{0}$. The solution to this equation is given by
\begin{align}
    \label{eq:reg2sol}
    \hat{\phi}(y) = \frac{1}{y^{4}}\frac{\beta}{40 \gamma^{2} \lambda} +\hat{\phi}_{0}\,.
\end{align}
The trajectory of $\hat \phi$ follows Eq.~(\ref{eq:reg2sol}) until $y \sim 1$ at which point the contribution from $\beta \hat{h}^{2}/2$ in Eq.~(\ref{eq:mastereq}) turns on and, to a good approximation, cancels the $\kappa^{2} \hat{\phi}$ term. 
At this point, all source terms in Eq.~(\ref{eq:reg2sol}) are negligible and the evolution is dictated by $\hat{\phi}''(y) = 0$. 
This has the solution
\begin{align}
    \label{eq:reg2sol2}
    \hat{\phi}(y) = \hat{\phi}'(1) (y-1) +  \hat{\phi}(1)\,,
\end{align}
where the initial conditions for the field and its velocity are obtained by matching the solution in Eq.~(\ref{eq:reg2sol}) at $y = 1$.
We find $\hat{\phi}'(1) = -\beta/(10 \gamma^{2} \lambda)$ and $\hat{\phi}(1) = \beta/(40 \gamma^{2}  \lambda) +\hat{\phi}_{0}$.
Then for small $y$, before beginning of oscillations, we obtain the asymptotic value of $\hat{\phi}(y \ll 1) = 5\beta/(40 \gamma^{2} \lambda) + \hat{\phi}_{0}$, where the first term is the amplitude of oscillations. The evolution of the scalar field for $\hat \phi_i = \hat \phi_0$ is displayed for a representative benchmark model in the left panel of Fig.~\ref{fig:Reg2_phi}.

With an estimate for the oscillation amplitude, $\hat \phi(y_{\rm osc}) \simeq 5\beta/(40 \gamma^{2} \lambda)$, we are now in a position to compute the $\phi$ relic abundance, following similar steps to those used earlier in Region I. 
In particular, we employ Eq.~(\ref{eq:phi-relic-density}) to estimate $\rho_{\phi,0}$\,, taking $y_f = y_{\rm osc}=\sqrt{\kappa/(3\gamma)}$, Eq.~(\ref{eq:yosc}), as the temperature corresponding to the onset of oscillations. 
Furthermore, we use $\rho(y_f) = \rho(y_{\rm osc}) \simeq \tfrac{1}{2} m_\phi^2 \phi(y_{\rm osc})^2$. 
Collecting the other inputs described around Eq.~(\ref{eq:phi-relic-density}), we obtain
\begin{align}
    \label{eq:ODMreg3}
    \Omega_{\phi,0}
    &\simeq  \frac{g_{*S}^0}{g_{*S}^{\rm osc}}\, \frac{y_0^3 \, \mu^4}{ \rho_{c,0}} \,
    \frac{3^{3/2}}{128} \, \frac{\beta^2 \, \kappa^{1/2}}{\gamma^{5/2}\, \lambda^2} \nonumber \\
    &\simeq 0.26 \left(\frac{\beta}{10^{-4}}\right)^{2} \left(\frac{\kappa}{4 \times 10^{-2}}\right)^{1/2}\,.
\end{align}
It should be noted that since the amplitude of oscillations is comparatively small for $\phi_{i} = \phi_{0}$, a much larger value of $\beta$ is needed to obtain the correct relic density. This initial condition is representative of an optimistic scenario for experimental detection of low mass scalars.

{\bf (ii)  $\hat {\phi}_{i} = 0$}: ~~ Next, we consider the initial condition $\hat{\phi}_{i} = 0$. It is far more straightforward to estimate the scalar relic density in this case. 
Thermal misalignment is again negligible, and in the earliest stages of its evolution the scalar is held near at its initial value by Hubble friction until the EWPT at $y_c$. 
At this point, the $\hat \phi$ VEV rapidly evolves to its zero temperature value $\hat {\phi}_0$, triggering VEV misalignment of the scalar, until eventually oscillations begin~(see Fig.~\ref{fig:Reg2_phi} right panel). 
The oscillation amplitude is therefore given by $\hat \phi(y_{\rm osc}) \simeq \hat{\phi}_{0} \simeq -\beta/(2 \lambda \kappa^{2})$. Thus the $\phi$ density parameter today is given by
\begin{align}
    \label{eq:ODMreg31}
    \Omega_{\phi,0}
    &\simeq  \frac{g_{*S}^0}{g_{*S}^{\rm osc}}\, \frac{y_0^3 \, \mu^4}{ \rho_{c,0}} \,
    \frac{3^{3/2}}{8} \, \frac{\gamma^{3/2} \beta^2 }{ \lambda^2  \, \kappa^{7/2}} \nonumber \\
    &\simeq 0.26 \left(\frac{\beta}{3 \times 10^{-10}}\right)^{2} \left(\frac{10^{-2}}{\kappa}\right)^{7/2}\,.
\end{align}

For this initial condition, we note the drastically lower values of $\beta$ required to get the correct relic density. Thus the initial condition $\phi_{i} = 0$ is representative of low mass scalar scenario that is challenging to probe experimentally.

\subsection{Summary of Results}

A summary of our results is presented in Fig.~\ref{fig:relic}, which shows contours leading to the observed relic abundance in the $m_\phi-A$ plane for the two choices of initial conditions, $\phi_i = \phi_0$ and $\phi_i = 0$. 
Both the numerical results from solving Eq.~(\ref{eq:mastereq}) and using Eq.~(\ref{eq:phi-relic-density}) (solid lines), as well as the analytic estimates, Eqs.~(\ref{eq:ODMreg1},\ref{eq:ODMreg3},\ref{eq:ODMreg31}) (dashed lines) are presented and agree well.

\section{Constraints}
\label{sec:results}

We now discuss the existing experimental constraints and future prospects for the ultra-light Higgs portal DM model. 
These bounds are shown in the $m_{\phi}-A$ plane and can be compared with our numerical and analytic predictions for the relic abundance lines ($\Omega_{\phi,0} = \Omega_{\rm DM} = 0.26$) in Fig.~\ref{fig:relic}. We now provide a brief overview for each of these bounds.
%

\subsection{Equivalence Principle and Inverse Square Law Tests}
The light scalar mixes with the Higgs and thus couples to ordinary matter,  leading to an additional Yukawa-like potential between nucleons. This additional force can be constrained by experiments searching for violations of the gravitational inverse square law (ISL) or tests of the equivalence principle (EP). 

For the ISL bounds, we follow the approach of 
Ref.~\cite{Piazza:2010ye}. 
The potential between two test bodies (labeled 1 and 2) is given by
\begin{align}
V(r)=-\frac{Gm_{1} m_{2}}{r}\left(1+\alpha_{1}  \alpha_{2} \, e^{-m_{\phi}r}\right)\,.
\end{align}
We will assume that the scalar has the same coupling to protons and neutrons, with the scalar-nucleon coupling given by
\begin{align}
    \label{eq:nucleoncoupling}
    g_{\phi N N} = g_{h N N} \,\frac{A\, v}{m^{2}_{h}} \simeq 10^{-3} \,\frac{A \, v}{m^{2}_{h}}\,,
\end{align}
where $g_{h N N} \simeq 10^{-3}$ is known with  
order one  uncertainty~\cite{Graham:2015ifn}.
The scalar coupling to the test body $i$ is given by
\begin{align}
\label{eq:coupling}
\frac{\alpha_{i}}{\sqrt{2}M_{\rm pl}} = \frac{d \ln m_{N}(\phi)}{d \phi} =  \frac{g_{\phi N N}}{m_{N}}\,.
\end{align}
We can then translate constraints reported on ISL tests. The constraints are typically reported on the product $\tilde{\alpha} = \alpha_{1} \alpha_{2}$, which in terms of our model parameters is given by
\begin{align}
    \label{eq:finconstraint}
    \tilde{\alpha} = \alpha_{1} \alpha_{2}  = 10^{-6}\, \frac{2 M^{2}_{\rm pl}}{m^{2}_{N}}\left(\frac{A \, v }{m^{2}_{h}}\right)^{2} = \left(\frac{A}{1.7\times 10^{-5} \, {\rm eV}}\right)^{2}.
\end{align}
The constraints from ISL tests from Refs.~\cite{Hoskins_1985,Yang_2012} are shown in Fig.~\ref{fig:relic}. As can be seen in the inset, the ISL tests are already beginning to probe interesting regions of the parameter space due to the resonance effects discussed in Sec.~\ref{sec:Reg2}.

Constraints from violation of EP are more involved. We follow Ref.~\cite{Graham:2015ifn} in the following. EP constraints involve measurements of differential accelerations in 
test bodies which have different charge-to-mass ratios, i.e., the ratio of the scalar coupling strength to
the mass of a given atom ($\Delta_{\phi H^{2}}$). The EP violation constraints on a mediator 
coupled to a $B-L$ charge obtained by measuring the differential acceleration between Beryllium (Be) and Aluminium (Al) were derived in Ref.~\cite{Wagner_2012}. Translating these bounds to 
our model, we can write
\begin{align}
    \label{eq:EP}
    A = 4.4 \times 10^{13} {\rm eV} \, g_{B-L} \, \sqrt{\frac{\Delta_{B-L}}{\Delta_{\Phi h^{2}}}}\,,
\end{align}
where $g_{B-L}$ is the the coupling strength of the scalar to $B-L$ charge and $\Delta_{B-L}$ is the differential charge-to-mass ratio between Be and Al. We use $\Delta_{B-L} = 0.037$ and $\Delta_{\phi H^{2}} = 4\times 10^{-4}$~\cite{Graham:2015ifn}. The EP constraints on the Higgs portal model are shown in Fig.~\ref{fig:relic}.

\subsection{Stellar Cooling Bounds}

For the Higgs portal, the stellar cooling bounds primarily arise from the effective coupling of $\phi$ to electrons \cite{Hardy:2016kme}. The dominant constraints on the model 
come from 
 red giant (RG) and horizontal branch (HB) stars. 
In red giant cores, the dominant energy loss mechanism is neutrino production. Any additional stronger energy loss mechanism will delay the onset of helium fusion in cores of RG stars. Since this is not observed, it implies a constraint on the Higgs-scalar mixing angle 
$\sin\theta \simeq  A v/m^{2}_{h} \leq 3\times 10^{-10}$ for $m_{\phi} < 2 ~ {\rm keV}$ \cite{Hardy:2016kme}. 
For HB stars, any additional strong energy loss mechanism causes the cores to contract and heat up, increasing the rate of helium fusion and shortening the lifetime of the stars, which is not observed. The combined constraints from RG and HB stars are shown in Fig. \ref{fig:relic}.

\subsection{Resonant Absorption in Molecules}

Resonant absorption of $\phi$ by polyatomic molecules has been proposed as a DM detection concept in Ref.~\cite{Arvanitaki:2017nhi}. 
The basic idea is to search for DM absorption via the subsequent transition of these molecules from the excited to ground state through the emission of photons. 
Concerning the Higgs portal model, this technique is most sensitive to the effective 
coupling of $\phi$ to electrons, which in terms of the coupling $A$ is given by
\begin{align}
    \label{eq:phie}
    g_{\phi e e} = \frac{A \, v}{m^{2}_{h}} y_{e} = \frac{A \, m_{e}}{m^{2}_{h}}.
\end{align}
The approach has the potential to probe scalar DM coupled to electrons in the mass range $ 0.2 \, {\rm eV} \lesssim m_{\phi} \lesssim 20 \, {\rm eV}$. In the parameterization employed in \cite{Arvanitaki:2017nhi}, the constraint is given by $d_{m_{e}} \lesssim 10^{5} $ where $d_{m_{e}} = \sqrt{2} g_{\phi e e} M_{\rm pl}/m_e \simeq  2.2 \times 10^{5} \times (A/{\rm eV})$.

We show projections corresponding to a phase II `bulk' (BII) and `stack' (SII) experimental configurations in Fig. \ref{fig:relic}; for more details we refer the reader to Ref.~\cite{Arvanitaki:2017nhi}.

\subsection{Extragalactic Background Light and Reionization Constraints}

For larger masses and couplings, the scalar $\phi$ can decay into photons on cosmological time scales. 
The decay rate of the scalar to photons is given by~\cite{Fradette:2018hhl}
\begin{align}
    \label{eq:gamma}
    \Gamma_{\phi \gamma \gamma} = \frac{\theta^{2} \alpha^{2}}{256 \pi ^{2}} \frac{m^{3}_{\phi}}{v^{2}} |C|^{2}
\end{align}
where $\theta \simeq A v/ m^{2}_{h}$ and the factor $C\simeq 50/27$ accounts for the charged particles running in the loop. 
Decay of scalars  would contribute to the extragalactic background light (EBL). EBL constraints have been calculated in \cite{Cadamuro:2011fd, Arias:2012az} and recast for a Higgs portal in \cite{Flacke:2016szy}. Note that these constrains are calculated assuming a given present day DM density $Y_{\phi} = n_{\phi}/s_{\gamma}$, where $s_{\gamma,0}$ is the entropy density of photons today. The actual parameter constrained is $m_{\phi} n_{\phi} \Gamma_{\phi}$ as a function of $m_{\phi}$. We rescale our constraints assuming $\Omega_{\phi} = \Omega_{DM} \simeq 0.26$ i.e. $Y_{\phi} = 0.26 \frac{\rho_{c}}{s_{\gamma,0}} m_{\phi}$, where $\rho_{c}$ is the critical energy density of universe today.

\subsection{X-ray Constraints}

For $m_\phi \gtrsim $ few keV, the scalar decay to photons on cosmological time scales is constrained by the HEAO~\cite{Gruber:1999yr} and INTEGRAL~\cite{Bouchet:2008rp} satellites. The data from these satellites was translated into lifetime of scalar decaying into photons in \cite{Essig:2013goa} and we use translate these constraints on the lifetime of our scalar $\tau_{\phi} = \Gamma^{-1}_{\phi \gamma \gamma}$ given by Eq.~(\ref{eq:gamma}).

\subsection{Other Probes}

There are other constraints that we have not shown because they lie outside the shown mass range, or lie in the forbidden region $A > \sqrt{2\lambda} m_{\phi}$. These include atomic and nuclear clocks, atomic interferometers, mechanical resonators, superradiance etc. A comprehensive summary of all constraints can be found in \cite{Antypas:2022asj}.

\section{Conclusions}
\label{sec:conclusion}

In this work we have revisited the cosmology of a light scalar feebly coupled to the SM through the super-renormalizable Higgs portal. This is among the most minimal UV complete extensions of the SM and is described by only two additional parameters, the scalar mass $m_\phi$ and its dimensionful coupling to the Higgs $A$. The scalar field in this model can naturally be light due to its  feeble super-renormalizable coupling. 
Despite being minimal, the model presents a rich cosmology, with the scalar field experiencing a non-trivial dynamical evolution during the radiation era before and/or during scalar oscillations. This generates two additional dynamical sources of misalignment, which we have referred to as thermal misalignment and VEV misalignment, beyond that from its initial field value at the end of inflation. We have studied the evolution of the scalar field both numerically and analytically over a broad range of scalar masses and couplings, delineating the parameters in the $m_\phi-A$ plane that are consistent with the observed DM relic density. We have also investigated two qualitatively distinct choices of initial scalar field values in order to discern the role played by initial conditions.  Our main results are presented in Fig.~\ref{fig:relic}.  

For large masses, the scalar field evolution is mainly governed by thermal misalignment. The finite-temperature contribution to the effective potential drives the scalar towards large field values at high temperatures, dynamically generating misalignment. 
Provided the initial scalar field value is smaller in magnitude than its displacement caused by thermal misalignment, the scalar relic density in this higher mass range is insensitive to the initial conditions. The scalar abundance today is then tightly controlled by the model parameters $m_\phi$ and $A$. Therefore, parameters in the mass-coupling plane yielding the observed relic abundance then provide a robust target that can be compared against experimental searches for ultra-light scalar DM. Furthermore, the relic density line in this mass range can be viewed as an upper bound on the scalar-Higgs coupling assuming a standard cosmology with a high enough reheat temperature, as any larger coupling would overclose the universe.
These features are quite similar in character to those of the popular WIMP DM scenario, in which the relic abundance is insensitive to initial conditions and UV dynamics,  instead  being set by the WIMP mass and its couplings to SM particles through the mechanism of thermal freeze-out. The insensitivity to initial conditions in the high scalar mass range is evident from Fig.~\ref{fig:relic}.  

For lower scalar masses, the impact of the EWPT on the scalar field evolution is important. The phase transition leads to a rapid shift in the scalar field VEV towards its zero-temperature VEV, generating VEV misalignment. However, in this regime the oscillation amplitude and corresponding relic density is also sensitive to the initial conditions and depends on whether the scalar field is initially in close proximity to its zero temperature minimum or not. In the former case, which can be naturally realized for a long period of inflation with a low inflationary Hubble parameter (smaller than the electroweak scale $v$), relatively large couplings are needed to significantly displace the field from its minimum and produce the observed DM abundance, thus offering more promising detection prospects. Instead, if the field is initially distant from its zero temperature VEV, the required coupling can be extremely small while still leading to a consistent DM cosmology.

In the intermediate mass range we have observed a novel forced resonance phenomena resulting from a competition between thermal misalignment and VEV misalignment. This causes enhancements or suppressions in the oscillation amplitude depending on the scalar mass, thus requiring smaller or larger couplings, respectively, to achieve the correct DM abundance. This is reflected by a series of peaks and valleys in the DM relic density contours as the scalar mass varies; see Fig.~\ref{fig:relic}.  In the future, it would be interesting to more accurately model the EWPT using lattice studies, which could impact the detailed character of the forced resonance phenomena.

We have also compared our relic density predictions with constraints and projections from a variety of terrestrial and astrophysical probes, including the  equivalence principle and inverse square law tests, stellar cooling, resonant molecular absorption, and observations of extra-galactic background light and diffuse X-ray backgrounds. While some parts of the parameter space are starting to be explored, new experimental ideas are needed to probe the cosmologically motivated regions of parameter space.  We hope our results will provide the impetus for new creative approaches to ultralight scalar DM detection.

\acknowledgments
We thank Daniel Boyanovsky, Raymond Co, Clara Murgui, Ryan Plestid, Michael Ratz, and Erwin Tanin for helpful discussions.  The work of B.B., A.G., and M.R. is supported by the U.S.~Department of Energy under grant No. DE–SC0007914. M.R. thanks ICTP, Trieste for hospitality while this work was being completed.

\appendix

\section{Thermal masses}
\label{sec:thermal-masses}
In this appendix we discuss the thermal masses $m_i^2(\phi,h,T)$ entering in the finite temperature potential, Eq.~(\ref{eq:VT}). 
As discussed in Section~\ref{sec:Veff}, we employ the daisy resummation scheme of Ref.~\cite{Parwani:1991gq}, writing
\begin{align}
    m_i^2(\phi,h,T) = m^2_{0,i}(\phi,h) + \Pi_i(T)\,.
\end{align}
We keep the leading contributions in the high temperature expansion for the self-energies, using the results presented in Ref.~\cite{Espinosa:1992kf}.

For the Higgs and Nambu-Goldstone scalars, these are  
\begin{align}
m_{0,h}^2(\phi,h) & = -\mu^2  +  3 \,\lambda \, h^2 + A \, \phi\,, \\
m_{0,\chi}^2(\phi,h) & = -\mu^2  +  \lambda \, h^2 + A \, \phi\,, \\
\Pi_{h}(T) & =\Pi_{\chi}(T)  = T^2\left( \frac{3}{16}g^2 + \frac{1}{16}g'^2 + \frac{1}{4}y_t^2 + \frac{1}{2}\lambda^2    \right)\,,
\end{align}
where $\mu^2$, $\lambda$, $A$ are parameters entering in the scalar potential, Eq.~(\ref{eq:potential}), and $g$, $g'$, and $y_t$ are the $SU(2)_L$ gauge coupling, $U(1)_Y$ gauge coupling, and top Yukawa coupling, respectively. 

In gauge boson sector, the squared masses for the transverse components are simply
\begin{align}
\label{mWZf}
m^2_{W_T}(h) = m_{0,W}^2(h) & = \frac{1}{4 }g^2 h^2\,, \\
m^2_{Z_T}(h) = m_{0,Z}^2(h) & = \frac{1}{4}(g^2+g'^2) h^2\,.
\end{align}
The longitudinal gauge boson components 
receive a self-energy correction, expressed in the gauge basis as $\Pi^L_{GB}(T)  = \frac{11}{6}T^2  {\rm diag}(g^2,g^2,g^2,g'^2)$. The thermal masses are thus given by the eigenvalues of the squared mass matrix 
\begin{align}
\left( 
\begin{array}{cccc}
\tfrac{1}{4}g^2h^2 & 0 & 0 & 0  \\
0 & \tfrac{1}{4} g^2h^2 & 0 & 0  \\
0 & 0 & \tfrac{1}{4} g^2h^2 & -\tfrac{1}{4} g g' h^2  \\
0 & 0 & -\tfrac{1}{4} g g' h^2 & \tfrac{1}{4} g'^2h^2  
\end{array}
\right)+ 
\left( 
\begin{array}{cccc}
\tfrac{11}{6}g^2 T^2 & 0 & 0 & 0  \\
0 & \tfrac{11}{6}g^2 T^2 & 0 & 0  \\
0 & 0 & \tfrac{11}{6}g^2 T^2 & 0  \\
0 & 0 & 0 & \tfrac{11}{6}g'^2 T^2  
\end{array}
\right).
\end{align}

Finally, for the top quark we have
\begin{align}
\label{mWZf}
m^2_{t}(h) = m_{0,t}^2(h) & = \frac{1}{2}y_t^2 h^2\,.
\end{align}

For completeness, we also collect here the expressions for the dimensionless arguments  
$\eta_{i}(\hat \phi, \hat h, y) \equiv m^{2}_{i}(\phi,h,T)/T^{2}$ of the $J_{B,F}$ functions. Using the results above and the dimensionless parameters defined in Eq.~(\ref{eq:dimless}), we obtain 
\begin{align}
    \label{eq:etas}
    \eta_{h} &= \frac{1}{y^{2}}\left[3 \lambda \hat{h}^{2} - (1-\beta \hat{\phi})\right] + \left(  \frac{3}{16}g^{2}+\frac{1}{16}g'^{2}+\frac{1}{4}y^{2}_{t} +\frac{1}{2}\lambda\right),\\
    \eta_{\chi} &= \frac{1}{y^{2}}\left[ \lambda \hat{h}^{2} - (1-\beta \hat{\phi})\right] + \left(  \frac{3}{16}g^{2}+\frac{1}{16}g'^{2}+\frac{1}{4}y^{2}_{t} +\frac{1}{2}\lambda\right),\\
    \eta_{W_{T}} &= \frac{g^{2} \hat{h}^{2}}{4 y^{2}}, \\
    \eta_{Z_{T}} &= \frac{(g^{2} + g'^{2} )\hat{h}^{2}}{4 y^{2}},\\
    \eta_{W_{L}} &= \frac{g^{2} \hat{h}^{2}}{4 y^{2}}  + \frac{11}{6}g^{2}, \\
    \eta_{Z_{L}} &= \frac{1}{2\,y^2} \left[\left(g^2+g'^2\right) \left(\tfrac{1}{4}\hat{h}^2+\tfrac{11}{6} y^2\right)
    +\sqrt{\left(g^2-g'^2\right)^2 \left(\tfrac{1}{4}\hat{h}^2+\tfrac{11}{6} y^2\right)^2 + \tfrac{1}{4} g^2 g'^2 {\hat h}^4~}\,\right], \\
    \eta_{A_{L}} &= \frac{1}{2\,y^2} \left[\left(g^2+g'^2\right) \left(\tfrac{1}{4}\hat{h}^2+\tfrac{11}{6} y^2\right)
    -\sqrt{\left(g^2-g'^2\right)^2 \left(\tfrac{1}{4}\hat{h}^2+\tfrac{11}{6} y^2\right)^2 + \tfrac{1}{4} g^2 g'^2 {\hat h}^4~}\,\right], \\
    \eta_{t} &= \frac{y_t^2 \hat h^2}{2 y^2} .
\end{align}

\bibliography{Higgs-TM}

\begin{thebibliography}{44}%
\makeatletter
\providecommand \@ifxundefined [1]{%
 \@ifx{#1\undefined}
}%
\providecommand \@ifnum [1]{%
 \ifnum #1\expandafter \@firstoftwo
 \else \expandafter \@secondoftwo
 \fi
}%
\providecommand \@ifx [1]{%
 \ifx #1\expandafter \@firstoftwo
 \else \expandafter \@secondoftwo
 \fi
}%
\providecommand \natexlab [1]{#1}%
\providecommand \enquote  [1]{``#1''}%
\providecommand \bibnamefont  [1]{#1}%
\providecommand \bibfnamefont [1]{#1}%
\providecommand \citenamefont [1]{#1}%
\providecommand \href@noop [0]{\@secondoftwo}%
\providecommand \href [0]{\begingroup \@sanitize@url \@href}%
\providecommand \@href[1]{\@@startlink{#1}\@@href}%
\providecommand \@@href[1]{\endgroup#1\@@endlink}%
\providecommand \@sanitize@url [0]{\catcode `\\12\catcode `\$12\catcode
  `\&12\catcode `\#12\catcode `\^12\catcode `\_12\catcode `\%12\relax}%
\providecommand \@@startlink[1]{}%
\providecommand \@@endlink[0]{}%
\providecommand \url  [0]{\begingroup\@sanitize@url \@url }%
\providecommand \@url [1]{\endgroup\@href {#1}{\urlprefix }}%
\providecommand \urlprefix  [0]{URL }%
\providecommand \Eprint [0]{\href }%
\providecommand \doibase [0]{http://dx.doi.org/}%
\providecommand \selectlanguage [0]{\@gobble}%
\providecommand \bibinfo  [0]{\@secondoftwo}%
\providecommand \bibfield  [0]{\@secondoftwo}%
\providecommand \translation [1]{[#1]}%
\providecommand \BibitemOpen [0]{}%
\providecommand \bibitemStop [0]{}%
\providecommand \bibitemNoStop [0]{.\EOS\space}%
\providecommand \EOS [0]{\spacefactor3000\relax}%
\providecommand \BibitemShut  [1]{\csname bibitem#1\endcsname}%
\let\auto@bib@innerbib\@empty
\bibitem [{\citenamefont {Aghanim}\ \emph {et~al.}(2020)\citenamefont {Aghanim}
  \emph {et~al.}}]{Planck:2018vyg}%
  \BibitemOpen
  \bibfield  {author} {\bibinfo {author} {\bibfnamefont {N.}~\bibnamefont
  {Aghanim}} \emph {et~al.} (\bibinfo {collaboration} {Planck}),\ }\href
  {\doibase 10.1051/0004-6361/201833910} {\bibfield  {journal} {\bibinfo
  {journal} {Astron. Astrophys.}\ }\textbf {\bibinfo {volume} {641}},\ \bibinfo
  {pages} {A6} (\bibinfo {year} {2020})},\ \Eprint
  {http://arxiv.org/abs/1807.06209} {arXiv:1807.06209 [astro-ph.CO]}
  \BibitemShut {NoStop}%
\bibitem [{\citenamefont {Antypas}\ \emph {et~al.}(2022)\citenamefont {Antypas}
  \emph {et~al.}}]{Antypas:2022asj}%
  \BibitemOpen
  \bibfield  {author} {\bibinfo {author} {\bibfnamefont {D.}~\bibnamefont
  {Antypas}} \emph {et~al.},\ }\href@noop {} {\  (\bibinfo {year} {2022})},\
  \Eprint {http://arxiv.org/abs/2203.14915} {arXiv:2203.14915 [hep-ex]}
  \BibitemShut {NoStop}%
\bibitem [{\citenamefont {Preskill}\ \emph {et~al.}(1983)\citenamefont
  {Preskill}, \citenamefont {Wise},\ and\ \citenamefont
  {Wilczek}}]{Preskill:1982cy}%
  \BibitemOpen
  \bibfield  {author} {\bibinfo {author} {\bibfnamefont {J.}~\bibnamefont
  {Preskill}}, \bibinfo {author} {\bibfnamefont {M.~B.}\ \bibnamefont {Wise}},
  \ and\ \bibinfo {author} {\bibfnamefont {F.}~\bibnamefont {Wilczek}},\ }\href
  {\doibase 10.1016/0370-2693(83)90637-8} {\bibfield  {journal} {\bibinfo
  {journal} {Phys. Lett. B}\ }\textbf {\bibinfo {volume} {120}},\ \bibinfo
  {pages} {127} (\bibinfo {year} {1983})}\BibitemShut {NoStop}%
\bibitem [{\citenamefont {Abbott}\ and\ \citenamefont
  {Sikivie}(1983)}]{Abbott:1982af}%
  \BibitemOpen
  \bibfield  {author} {\bibinfo {author} {\bibfnamefont {L.~F.}\ \bibnamefont
  {Abbott}}\ and\ \bibinfo {author} {\bibfnamefont {P.}~\bibnamefont
  {Sikivie}},\ }\href {\doibase 10.1016/0370-2693(83)90638-X} {\bibfield
  {journal} {\bibinfo  {journal} {Phys. Lett. B}\ }\textbf {\bibinfo {volume}
  {120}},\ \bibinfo {pages} {133} (\bibinfo {year} {1983})}\BibitemShut
  {NoStop}%
\bibitem [{\citenamefont {Dine}\ and\ \citenamefont
  {Fischler}(1983)}]{Dine:1982ah}%
  \BibitemOpen
  \bibfield  {author} {\bibinfo {author} {\bibfnamefont {M.}~\bibnamefont
  {Dine}}\ and\ \bibinfo {author} {\bibfnamefont {W.}~\bibnamefont
  {Fischler}},\ }\href {\doibase 10.1016/0370-2693(83)90639-1} {\bibfield
  {journal} {\bibinfo  {journal} {Phys. Lett. B}\ }\textbf {\bibinfo {volume}
  {120}},\ \bibinfo {pages} {137} (\bibinfo {year} {1983})}\BibitemShut
  {NoStop}%
\bibitem [{\citenamefont {Piazza}\ and\ \citenamefont
  {Pospelov}(2010)}]{Piazza:2010ye}%
  \BibitemOpen
  \bibfield  {author} {\bibinfo {author} {\bibfnamefont {F.}~\bibnamefont
  {Piazza}}\ and\ \bibinfo {author} {\bibfnamefont {M.}~\bibnamefont
  {Pospelov}},\ }\href {\doibase 10.1103/PhysRevD.82.043533} {\bibfield
  {journal} {\bibinfo  {journal} {Phys. Rev. D}\ }\textbf {\bibinfo {volume}
  {82}},\ \bibinfo {pages} {043533} (\bibinfo {year} {2010})},\ \Eprint
  {http://arxiv.org/abs/1003.2313} {arXiv:1003.2313 [hep-ph]} \BibitemShut
  {NoStop}%
\bibitem [{\citenamefont {Graham}\ \emph {et~al.}(2016)\citenamefont {Graham},
  \citenamefont {Kaplan}, \citenamefont {Mardon}, \citenamefont {Rajendran},\
  and\ \citenamefont {Terrano}}]{Graham:2015ifn}%
  \BibitemOpen
  \bibfield  {author} {\bibinfo {author} {\bibfnamefont {P.~W.}\ \bibnamefont
  {Graham}}, \bibinfo {author} {\bibfnamefont {D.~E.}\ \bibnamefont {Kaplan}},
  \bibinfo {author} {\bibfnamefont {J.}~\bibnamefont {Mardon}}, \bibinfo
  {author} {\bibfnamefont {S.}~\bibnamefont {Rajendran}}, \ and\ \bibinfo
  {author} {\bibfnamefont {W.~A.}\ \bibnamefont {Terrano}},\ }\href {\doibase
  10.1103/PhysRevD.93.075029} {\bibfield  {journal} {\bibinfo  {journal} {Phys.
  Rev. D}\ }\textbf {\bibinfo {volume} {93}},\ \bibinfo {pages} {075029}
  (\bibinfo {year} {2016})},\ \Eprint {http://arxiv.org/abs/1512.06165}
  {arXiv:1512.06165 [hep-ph]} \BibitemShut {NoStop}%
\bibitem [{\citenamefont {Arvanitaki}\ \emph {et~al.}(2018)\citenamefont
  {Arvanitaki}, \citenamefont {Dimopoulos},\ and\ \citenamefont
  {Van~Tilburg}}]{Arvanitaki:2017nhi}%
  \BibitemOpen
  \bibfield  {author} {\bibinfo {author} {\bibfnamefont {A.}~\bibnamefont
  {Arvanitaki}}, \bibinfo {author} {\bibfnamefont {S.}~\bibnamefont
  {Dimopoulos}}, \ and\ \bibinfo {author} {\bibfnamefont {K.}~\bibnamefont
  {Van~Tilburg}},\ }\href {\doibase 10.1103/PhysRevX.8.041001} {\bibfield
  {journal} {\bibinfo  {journal} {Phys. Rev. X}\ }\textbf {\bibinfo {volume}
  {8}},\ \bibinfo {pages} {041001} (\bibinfo {year} {2018})},\ \Eprint
  {http://arxiv.org/abs/1709.05354} {arXiv:1709.05354 [hep-ph]} \BibitemShut
  {NoStop}%
\bibitem [{\citenamefont {Batell}\ and\ \citenamefont
  {Ghalsasi}(2023)}]{Batell:2021ofv}%
  \BibitemOpen
  \bibfield  {author} {\bibinfo {author} {\bibfnamefont {B.}~\bibnamefont
  {Batell}}\ and\ \bibinfo {author} {\bibfnamefont {A.}~\bibnamefont
  {Ghalsasi}},\ }\href {\doibase 10.1103/PhysRevD.107.L091701} {\bibfield
  {journal} {\bibinfo  {journal} {Phys. Rev. D}\ }\textbf {\bibinfo {volume}
  {107}},\ \bibinfo {pages} {L091701} (\bibinfo {year} {2023})},\ \Eprint
  {http://arxiv.org/abs/2109.04476} {arXiv:2109.04476 [hep-ph]} \BibitemShut
  {NoStop}%
\bibitem [{\citenamefont {Buchmuller}\ \emph {et~al.}(2003)\citenamefont
  {Buchmuller}, \citenamefont {Hamaguchi},\ and\ \citenamefont
  {Ratz}}]{Buchmuller:2003is}%
  \BibitemOpen
  \bibfield  {author} {\bibinfo {author} {\bibfnamefont {W.}~\bibnamefont
  {Buchmuller}}, \bibinfo {author} {\bibfnamefont {K.}~\bibnamefont
  {Hamaguchi}}, \ and\ \bibinfo {author} {\bibfnamefont {M.}~\bibnamefont
  {Ratz}},\ }\href {\doibase 10.1016/j.physletb.2003.09.017} {\bibfield
  {journal} {\bibinfo  {journal} {Phys. Lett. B}\ }\textbf {\bibinfo {volume}
  {574}},\ \bibinfo {pages} {156} (\bibinfo {year} {2003})},\ \Eprint
  {http://arxiv.org/abs/hep-ph/0307181} {arXiv:hep-ph/0307181} \BibitemShut
  {NoStop}%
\bibitem [{\citenamefont {Fardon}\ \emph {et~al.}(2004)\citenamefont {Fardon},
  \citenamefont {Nelson},\ and\ \citenamefont {Weiner}}]{Fardon:2003eh}%
  \BibitemOpen
  \bibfield  {author} {\bibinfo {author} {\bibfnamefont {R.}~\bibnamefont
  {Fardon}}, \bibinfo {author} {\bibfnamefont {A.~E.}\ \bibnamefont {Nelson}},
  \ and\ \bibinfo {author} {\bibfnamefont {N.}~\bibnamefont {Weiner}},\ }\href
  {\doibase 10.1088/1475-7516/2004/10/005} {\bibfield  {journal} {\bibinfo
  {journal} {JCAP}\ }\textbf {\bibinfo {volume} {10}},\ \bibinfo {pages} {005}
  (\bibinfo {year} {2004})},\ \Eprint {http://arxiv.org/abs/astro-ph/0309800}
  {arXiv:astro-ph/0309800} \BibitemShut {NoStop}%
\bibitem [{\citenamefont {Buchmuller}\ \emph {et~al.}(2004)\citenamefont
  {Buchmuller}, \citenamefont {Hamaguchi}, \citenamefont {Lebedev},\ and\
  \citenamefont {Ratz}}]{Buchmuller:2004xr}%
  \BibitemOpen
  \bibfield  {author} {\bibinfo {author} {\bibfnamefont {W.}~\bibnamefont
  {Buchmuller}}, \bibinfo {author} {\bibfnamefont {K.}~\bibnamefont
  {Hamaguchi}}, \bibinfo {author} {\bibfnamefont {O.}~\bibnamefont {Lebedev}},
  \ and\ \bibinfo {author} {\bibfnamefont {M.}~\bibnamefont {Ratz}},\ }\href
  {\doibase 10.1016/j.nuclphysb.2004.08.031} {\bibfield  {journal} {\bibinfo
  {journal} {Nucl. Phys. B}\ }\textbf {\bibinfo {volume} {699}},\ \bibinfo
  {pages} {292} (\bibinfo {year} {2004})},\ \Eprint
  {http://arxiv.org/abs/hep-th/0404168} {arXiv:hep-th/0404168} \BibitemShut
  {NoStop}%
\bibitem [{\citenamefont {Moroi}\ \emph {et~al.}(2013)\citenamefont {Moroi},
  \citenamefont {Mukaida}, \citenamefont {Nakayama},\ and\ \citenamefont
  {Takimoto}}]{Moroi:2013tea}%
  \BibitemOpen
  \bibfield  {author} {\bibinfo {author} {\bibfnamefont {T.}~\bibnamefont
  {Moroi}}, \bibinfo {author} {\bibfnamefont {K.}~\bibnamefont {Mukaida}},
  \bibinfo {author} {\bibfnamefont {K.}~\bibnamefont {Nakayama}}, \ and\
  \bibinfo {author} {\bibfnamefont {M.}~\bibnamefont {Takimoto}},\ }\href
  {\doibase 10.1007/JHEP06(2013)040} {\bibfield  {journal} {\bibinfo  {journal}
  {JHEP}\ }\textbf {\bibinfo {volume} {06}},\ \bibinfo {pages} {040} (\bibinfo
  {year} {2013})},\ \Eprint {http://arxiv.org/abs/1304.6597} {arXiv:1304.6597
  [hep-ph]} \BibitemShut {NoStop}%
\bibitem [{\citenamefont {Lillard}\ \emph {et~al.}(2018)\citenamefont
  {Lillard}, \citenamefont {Ratz}, \citenamefont {Tait},\ and\ \citenamefont
  {Trojanowski}}]{Lillard:2018zts}%
  \BibitemOpen
  \bibfield  {author} {\bibinfo {author} {\bibfnamefont {B.}~\bibnamefont
  {Lillard}}, \bibinfo {author} {\bibfnamefont {M.}~\bibnamefont {Ratz}},
  \bibinfo {author} {\bibfnamefont {T.}~\bibnamefont {Tait}, \bibfnamefont
  {M.~P.}}, \ and\ \bibinfo {author} {\bibfnamefont {S.}~\bibnamefont
  {Trojanowski}},\ }\href {\doibase 10.1088/1475-7516/2018/07/056} {\bibfield
  {journal} {\bibinfo  {journal} {JCAP}\ }\textbf {\bibinfo {volume} {07}},\
  \bibinfo {pages} {056} (\bibinfo {year} {2018})},\ \Eprint
  {http://arxiv.org/abs/1804.03662} {arXiv:1804.03662 [hep-ph]} \BibitemShut
  {NoStop}%
\bibitem [{\citenamefont {Chun}(2022)}]{Chun:2021uwr}%
  \BibitemOpen
  \bibfield  {author} {\bibinfo {author} {\bibfnamefont {E.~J.}\ \bibnamefont
  {Chun}},\ }\href {\doibase 10.1016/j.physletb.2022.136880} {\bibfield
  {journal} {\bibinfo  {journal} {Phys. Lett. B}\ }\textbf {\bibinfo {volume}
  {825}},\ \bibinfo {pages} {136880} (\bibinfo {year} {2022})},\ \Eprint
  {http://arxiv.org/abs/2109.07423} {arXiv:2109.07423 [hep-ph]} \BibitemShut
  {NoStop}%
\bibitem [{\citenamefont {Brzeminski}\ \emph {et~al.}(2021)\citenamefont
  {Brzeminski}, \citenamefont {Chacko}, \citenamefont {Dev},\ and\
  \citenamefont {Hook}}]{Brzeminski:2020uhm}%
  \BibitemOpen
  \bibfield  {author} {\bibinfo {author} {\bibfnamefont {D.}~\bibnamefont
  {Brzeminski}}, \bibinfo {author} {\bibfnamefont {Z.}~\bibnamefont {Chacko}},
  \bibinfo {author} {\bibfnamefont {A.}~\bibnamefont {Dev}}, \ and\ \bibinfo
  {author} {\bibfnamefont {A.}~\bibnamefont {Hook}},\ }\href {\doibase
  10.1103/PhysRevD.104.075019} {\bibfield  {journal} {\bibinfo  {journal}
  {Phys. Rev. D}\ }\textbf {\bibinfo {volume} {104}},\ \bibinfo {pages}
  {075019} (\bibinfo {year} {2021})},\ \Eprint
  {http://arxiv.org/abs/2012.02787} {arXiv:2012.02787 [hep-ph]} \BibitemShut
  {NoStop}%
\bibitem [{\citenamefont {Croon}\ \emph {et~al.}(2022)\citenamefont {Croon},
  \citenamefont {Davoudiasl},\ and\ \citenamefont {Houtz}}]{Croon:2022gwq}%
  \BibitemOpen
  \bibfield  {author} {\bibinfo {author} {\bibfnamefont {D.}~\bibnamefont
  {Croon}}, \bibinfo {author} {\bibfnamefont {H.}~\bibnamefont {Davoudiasl}}, \
  and\ \bibinfo {author} {\bibfnamefont {R.}~\bibnamefont {Houtz}},\ }\href
  {\doibase 10.1103/PhysRevD.106.035006} {\bibfield  {journal} {\bibinfo
  {journal} {Phys. Rev. D}\ }\textbf {\bibinfo {volume} {106}},\ \bibinfo
  {pages} {035006} (\bibinfo {year} {2022})},\ \Eprint
  {http://arxiv.org/abs/2204.07584} {arXiv:2204.07584 [hep-ph]} \BibitemShut
  {NoStop}%
\bibitem [{\citenamefont {Cheek}\ \emph {et~al.}(2023)\citenamefont {Cheek},
  \citenamefont {Osi\'nski}, \citenamefont {Roszkowski},\ and\ \citenamefont
  {Trojanowski}}]{Cheek:2022yof}%
  \BibitemOpen
  \bibfield  {author} {\bibinfo {author} {\bibfnamefont {A.}~\bibnamefont
  {Cheek}}, \bibinfo {author} {\bibfnamefont {J.~K.}\ \bibnamefont
  {Osi\'nski}}, \bibinfo {author} {\bibfnamefont {L.}~\bibnamefont
  {Roszkowski}}, \ and\ \bibinfo {author} {\bibfnamefont {S.}~\bibnamefont
  {Trojanowski}},\ }\href {\doibase 10.1007/JHEP03(2023)149} {\bibfield
  {journal} {\bibinfo  {journal} {JHEP}\ }\textbf {\bibinfo {volume} {03}},\
  \bibinfo {pages} {149} (\bibinfo {year} {2023})},\ \Eprint
  {http://arxiv.org/abs/2211.02057} {arXiv:2211.02057 [hep-ph]} \BibitemShut
  {NoStop}%
\bibitem [{\citenamefont {Arkani-Hamed}\ \emph {et~al.}(2021)\citenamefont
  {Arkani-Hamed}, \citenamefont {D'Agnolo},\ and\ \citenamefont
  {Kim}}]{Arkani-Hamed:2020yna}%
  \BibitemOpen
  \bibfield  {author} {\bibinfo {author} {\bibfnamefont {N.}~\bibnamefont
  {Arkani-Hamed}}, \bibinfo {author} {\bibfnamefont {R.~T.}\ \bibnamefont
  {D'Agnolo}}, \ and\ \bibinfo {author} {\bibfnamefont {H.~D.}\ \bibnamefont
  {Kim}},\ }\href {\doibase 10.1103/PhysRevD.104.095014} {\bibfield  {journal}
  {\bibinfo  {journal} {Phys. Rev. D}\ }\textbf {\bibinfo {volume} {104}},\
  \bibinfo {pages} {095014} (\bibinfo {year} {2021})},\ \Eprint
  {http://arxiv.org/abs/2012.04652} {arXiv:2012.04652 [hep-ph]} \BibitemShut
  {NoStop}%
\bibitem [{\citenamefont {Coleman}\ and\ \citenamefont
  {Weinberg}(1973)}]{Coleman:1973jx}%
  \BibitemOpen
  \bibfield  {author} {\bibinfo {author} {\bibfnamefont {S.~R.}\ \bibnamefont
  {Coleman}}\ and\ \bibinfo {author} {\bibfnamefont {E.~J.}\ \bibnamefont
  {Weinberg}},\ }\href {\doibase 10.1103/PhysRevD.7.1888} {\bibfield  {journal}
  {\bibinfo  {journal} {Phys. Rev. D}\ }\textbf {\bibinfo {volume} {7}},\
  \bibinfo {pages} {1888} (\bibinfo {year} {1973})}\BibitemShut {NoStop}%
\bibitem [{\citenamefont {Dolan}\ and\ \citenamefont
  {Jackiw}(1974)}]{Dolan:1973qd}%
  \BibitemOpen
  \bibfield  {author} {\bibinfo {author} {\bibfnamefont {L.}~\bibnamefont
  {Dolan}}\ and\ \bibinfo {author} {\bibfnamefont {R.}~\bibnamefont {Jackiw}},\
  }\href {\doibase 10.1103/PhysRevD.9.3320} {\bibfield  {journal} {\bibinfo
  {journal} {Phys. Rev. D}\ }\textbf {\bibinfo {volume} {9}},\ \bibinfo {pages}
  {3320} (\bibinfo {year} {1974})}\BibitemShut {NoStop}%
\bibitem [{\citenamefont {Weinberg}(1974)}]{Weinberg:1974hy}%
  \BibitemOpen
  \bibfield  {author} {\bibinfo {author} {\bibfnamefont {S.}~\bibnamefont
  {Weinberg}},\ }\href {\doibase 10.1103/PhysRevD.9.3357} {\bibfield  {journal}
  {\bibinfo  {journal} {Phys. Rev. D}\ }\textbf {\bibinfo {volume} {9}},\
  \bibinfo {pages} {3357} (\bibinfo {year} {1974})}\BibitemShut {NoStop}%
\bibitem [{\citenamefont {Quiros}(1999)}]{Quiros:1999jp}%
  \BibitemOpen
  \bibfield  {author} {\bibinfo {author} {\bibfnamefont {M.}~\bibnamefont
  {Quiros}},\ }in\ \href@noop {} {\emph {\bibinfo {booktitle} {{ICTP Summer
  School in High-Energy Physics and Cosmology}}}}\ (\bibinfo {year} {1999})\
  \Eprint {http://arxiv.org/abs/hep-ph/9901312} {arXiv:hep-ph/9901312}
  \BibitemShut {NoStop}%
\bibitem [{\citenamefont {Parwani}(1992)}]{Parwani:1991gq}%
  \BibitemOpen
  \bibfield  {author} {\bibinfo {author} {\bibfnamefont {R.~R.}\ \bibnamefont
  {Parwani}},\ }\href {\doibase 10.1103/PhysRevD.45.4695} {\bibfield  {journal}
  {\bibinfo  {journal} {Phys. Rev. D}\ }\textbf {\bibinfo {volume} {45}},\
  \bibinfo {pages} {4695} (\bibinfo {year} {1992})},\ \bibinfo {note}
  {[Erratum: Phys.Rev.D 48, 5965 (1993)]},\ \Eprint
  {http://arxiv.org/abs/hep-ph/9204216} {arXiv:hep-ph/9204216} \BibitemShut
  {NoStop}%
\bibitem [{\citenamefont {Kajantie}\ \emph {et~al.}(1996)\citenamefont
  {Kajantie}, \citenamefont {Laine}, \citenamefont {Rummukainen},\ and\
  \citenamefont {Shaposhnikov}}]{Kajantie:1996mn}%
  \BibitemOpen
  \bibfield  {author} {\bibinfo {author} {\bibfnamefont {K.}~\bibnamefont
  {Kajantie}}, \bibinfo {author} {\bibfnamefont {M.}~\bibnamefont {Laine}},
  \bibinfo {author} {\bibfnamefont {K.}~\bibnamefont {Rummukainen}}, \ and\
  \bibinfo {author} {\bibfnamefont {M.~E.}\ \bibnamefont {Shaposhnikov}},\
  }\href {\doibase 10.1103/PhysRevLett.77.2887} {\bibfield  {journal} {\bibinfo
   {journal} {Phys. Rev. Lett.}\ }\textbf {\bibinfo {volume} {77}},\ \bibinfo
  {pages} {2887} (\bibinfo {year} {1996})},\ \Eprint
  {http://arxiv.org/abs/hep-ph/9605288} {arXiv:hep-ph/9605288} \BibitemShut
  {NoStop}%
\bibitem [{\citenamefont {D'Onofrio}\ and\ \citenamefont
  {Rummukainen}(2016)}]{DOnofrio:2015gop}%
  \BibitemOpen
  \bibfield  {author} {\bibinfo {author} {\bibfnamefont {M.}~\bibnamefont
  {D'Onofrio}}\ and\ \bibinfo {author} {\bibfnamefont {K.}~\bibnamefont
  {Rummukainen}},\ }\href {\doibase 10.1103/PhysRevD.93.025003} {\bibfield
  {journal} {\bibinfo  {journal} {Phys. Rev. D}\ }\textbf {\bibinfo {volume}
  {93}},\ \bibinfo {pages} {025003} (\bibinfo {year} {2016})},\ \Eprint
  {http://arxiv.org/abs/1508.07161} {arXiv:1508.07161 [hep-ph]} \BibitemShut
  {NoStop}%
\bibitem [{\citenamefont {Graham}\ and\ \citenamefont
  {Scherlis}(2018)}]{Graham:2018jyp}%
  \BibitemOpen
  \bibfield  {author} {\bibinfo {author} {\bibfnamefont {P.~W.}\ \bibnamefont
  {Graham}}\ and\ \bibinfo {author} {\bibfnamefont {A.}~\bibnamefont
  {Scherlis}},\ }\href {\doibase 10.1103/PhysRevD.98.035017} {\bibfield
  {journal} {\bibinfo  {journal} {Phys. Rev. D}\ }\textbf {\bibinfo {volume}
  {98}},\ \bibinfo {pages} {035017} (\bibinfo {year} {2018})},\ \Eprint
  {http://arxiv.org/abs/1805.07362} {arXiv:1805.07362 [hep-ph]} \BibitemShut
  {NoStop}%
\bibitem [{\citenamefont {Takahashi}\ \emph {et~al.}(2018)\citenamefont
  {Takahashi}, \citenamefont {Yin},\ and\ \citenamefont
  {Guth}}]{Takahashi:2018tdu}%
  \BibitemOpen
  \bibfield  {author} {\bibinfo {author} {\bibfnamefont {F.}~\bibnamefont
  {Takahashi}}, \bibinfo {author} {\bibfnamefont {W.}~\bibnamefont {Yin}}, \
  and\ \bibinfo {author} {\bibfnamefont {A.~H.}\ \bibnamefont {Guth}},\ }\href
  {\doibase 10.1103/PhysRevD.98.015042} {\bibfield  {journal} {\bibinfo
  {journal} {Phys. Rev. D}\ }\textbf {\bibinfo {volume} {98}},\ \bibinfo
  {pages} {015042} (\bibinfo {year} {2018})},\ \Eprint
  {http://arxiv.org/abs/1805.08763} {arXiv:1805.08763 [hep-ph]} \BibitemShut
  {NoStop}%
\bibitem [{\citenamefont {Tenkanen}(2019)}]{Tenkanen:2019aij}%
  \BibitemOpen
  \bibfield  {author} {\bibinfo {author} {\bibfnamefont {T.}~\bibnamefont
  {Tenkanen}},\ }\href {\doibase 10.1103/PhysRevLett.123.061302} {\bibfield
  {journal} {\bibinfo  {journal} {Phys. Rev. Lett.}\ }\textbf {\bibinfo
  {volume} {123}},\ \bibinfo {pages} {061302} (\bibinfo {year} {2019})},\
  \Eprint {http://arxiv.org/abs/1905.01214} {arXiv:1905.01214 [astro-ph.CO]}
  \BibitemShut {NoStop}%
\bibitem [{\citenamefont {Akrami}\ \emph {et~al.}(2020)\citenamefont {Akrami}
  \emph {et~al.}}]{Planck:2018jri}%
  \BibitemOpen
  \bibfield  {author} {\bibinfo {author} {\bibfnamefont {Y.}~\bibnamefont
  {Akrami}} \emph {et~al.} (\bibinfo {collaboration} {Planck}),\ }\href
  {\doibase 10.1051/0004-6361/201833887} {\bibfield  {journal} {\bibinfo
  {journal} {Astron. Astrophys.}\ }\textbf {\bibinfo {volume} {641}},\ \bibinfo
  {pages} {A10} (\bibinfo {year} {2020})},\ \Eprint
  {http://arxiv.org/abs/1807.06211} {arXiv:1807.06211 [astro-ph.CO]}
  \BibitemShut {NoStop}%
\bibitem [{\citenamefont {Yokoyama}\ and\ \citenamefont
  {Linde}(1999)}]{Yokoyama:1998ju}%
  \BibitemOpen
  \bibfield  {author} {\bibinfo {author} {\bibfnamefont {J.}~\bibnamefont
  {Yokoyama}}\ and\ \bibinfo {author} {\bibfnamefont {A.~D.}\ \bibnamefont
  {Linde}},\ }\href {\doibase 10.1103/PhysRevD.60.083509} {\bibfield  {journal}
  {\bibinfo  {journal} {Phys. Rev. D}\ }\textbf {\bibinfo {volume} {60}},\
  \bibinfo {pages} {083509} (\bibinfo {year} {1999})},\ \Eprint
  {http://arxiv.org/abs/hep-ph/9809409} {arXiv:hep-ph/9809409} \BibitemShut
  {NoStop}%
\bibitem [{\citenamefont {B\"odeker}\ and\ \citenamefont
  {Nienaber}(2022)}]{Bodeker:2022ihg}%
  \BibitemOpen
  \bibfield  {author} {\bibinfo {author} {\bibfnamefont {D.}~\bibnamefont
  {B\"odeker}}\ and\ \bibinfo {author} {\bibfnamefont {J.}~\bibnamefont
  {Nienaber}},\ }\href {\doibase 10.1103/PhysRevD.106.056016} {\bibfield
  {journal} {\bibinfo  {journal} {Phys. Rev. D}\ }\textbf {\bibinfo {volume}
  {106}},\ \bibinfo {pages} {056016} (\bibinfo {year} {2022})},\ \Eprint
  {http://arxiv.org/abs/2205.14166} {arXiv:2205.14166 [hep-ph]} \BibitemShut
  {NoStop}%
\bibitem [{\citenamefont {Hoskins}\ \emph {et~al.}(1985)\citenamefont
  {Hoskins}, \citenamefont {Newman}, \citenamefont {Spero},\ and\ \citenamefont
  {Schultz}}]{Hoskins_1985}%
  \BibitemOpen
  \bibfield  {author} {\bibinfo {author} {\bibfnamefont {J.~K.}\ \bibnamefont
  {Hoskins}}, \bibinfo {author} {\bibfnamefont {R.~D.}\ \bibnamefont {Newman}},
  \bibinfo {author} {\bibfnamefont {R.}~\bibnamefont {Spero}}, \ and\ \bibinfo
  {author} {\bibfnamefont {J.}~\bibnamefont {Schultz}},\ }\href {\doibase
  10.1103/PhysRevD.32.3084} {\bibfield  {journal} {\bibinfo  {journal} {Phys.
  Rev. D}\ }\textbf {\bibinfo {volume} {32}},\ \bibinfo {pages} {3084}
  (\bibinfo {year} {1985})}\BibitemShut {NoStop}%
\bibitem [{\citenamefont {Yang}\ \emph {et~al.}(2012)\citenamefont {Yang},
  \citenamefont {Zhan}, \citenamefont {Wang}, \citenamefont {Shao},
  \citenamefont {Tu}, \citenamefont {Tan},\ and\ \citenamefont
  {Luo}}]{Yang_2012}%
  \BibitemOpen
  \bibfield  {author} {\bibinfo {author} {\bibfnamefont {S.-Q.}\ \bibnamefont
  {Yang}}, \bibinfo {author} {\bibfnamefont {B.-F.}\ \bibnamefont {Zhan}},
  \bibinfo {author} {\bibfnamefont {Q.-L.}\ \bibnamefont {Wang}}, \bibinfo
  {author} {\bibfnamefont {C.-G.}\ \bibnamefont {Shao}}, \bibinfo {author}
  {\bibfnamefont {L.-C.}\ \bibnamefont {Tu}}, \bibinfo {author} {\bibfnamefont
  {W.-H.}\ \bibnamefont {Tan}}, \ and\ \bibinfo {author} {\bibfnamefont
  {J.}~\bibnamefont {Luo}},\ }\href {\doibase 10.1103/PhysRevLett.108.081101}
  {\bibfield  {journal} {\bibinfo  {journal} {Phys. Rev. Lett.}\ }\textbf
  {\bibinfo {volume} {108}},\ \bibinfo {pages} {081101} (\bibinfo {year}
  {2012})}\BibitemShut {NoStop}%
\bibitem [{\citenamefont {Wagner}\ \emph {et~al.}(2012)\citenamefont {Wagner},
  \citenamefont {Schlamminger}, \citenamefont {Gundlach},\ and\ \citenamefont
  {Adelberger}}]{Wagner_2012}%
  \BibitemOpen
  \bibfield  {author} {\bibinfo {author} {\bibfnamefont {T.~A.}\ \bibnamefont
  {Wagner}}, \bibinfo {author} {\bibfnamefont {S.}~\bibnamefont
  {Schlamminger}}, \bibinfo {author} {\bibfnamefont {J.~H.}\ \bibnamefont
  {Gundlach}}, \ and\ \bibinfo {author} {\bibfnamefont {E.~G.}\ \bibnamefont
  {Adelberger}},\ }\href {\doibase 10.1088/0264-9381/29/18/184002} {\bibfield
  {journal} {\bibinfo  {journal} {Classical and Quantum Gravity}\ }\textbf
  {\bibinfo {volume} {29}},\ \bibinfo {pages} {184002} (\bibinfo {year}
  {2012})}\BibitemShut {NoStop}%
\bibitem [{\citenamefont {Hardy}\ and\ \citenamefont
  {Lasenby}(2017)}]{Hardy:2016kme}%
  \BibitemOpen
  \bibfield  {author} {\bibinfo {author} {\bibfnamefont {E.}~\bibnamefont
  {Hardy}}\ and\ \bibinfo {author} {\bibfnamefont {R.}~\bibnamefont
  {Lasenby}},\ }\href {\doibase 10.1007/JHEP02(2017)033} {\bibfield  {journal}
  {\bibinfo  {journal} {JHEP}\ }\textbf {\bibinfo {volume} {02}},\ \bibinfo
  {pages} {033} (\bibinfo {year} {2017})},\ \Eprint
  {http://arxiv.org/abs/1611.05852} {arXiv:1611.05852 [hep-ph]} \BibitemShut
  {NoStop}%
\bibitem [{\citenamefont {Fradette}\ \emph {et~al.}(2019)\citenamefont
  {Fradette}, \citenamefont {Pospelov}, \citenamefont {Pradler},\ and\
  \citenamefont {Ritz}}]{Fradette:2018hhl}%
  \BibitemOpen
  \bibfield  {author} {\bibinfo {author} {\bibfnamefont {A.}~\bibnamefont
  {Fradette}}, \bibinfo {author} {\bibfnamefont {M.}~\bibnamefont {Pospelov}},
  \bibinfo {author} {\bibfnamefont {J.}~\bibnamefont {Pradler}}, \ and\
  \bibinfo {author} {\bibfnamefont {A.}~\bibnamefont {Ritz}},\ }\href {\doibase
  10.1103/PhysRevD.99.075004} {\bibfield  {journal} {\bibinfo  {journal} {Phys.
  Rev. D}\ }\textbf {\bibinfo {volume} {99}},\ \bibinfo {pages} {075004}
  (\bibinfo {year} {2019})},\ \Eprint {http://arxiv.org/abs/1812.07585}
  {arXiv:1812.07585 [hep-ph]} \BibitemShut {NoStop}%
\bibitem [{\citenamefont {Cadamuro}\ and\ \citenamefont
  {Redondo}(2012)}]{Cadamuro:2011fd}%
  \BibitemOpen
  \bibfield  {author} {\bibinfo {author} {\bibfnamefont {D.}~\bibnamefont
  {Cadamuro}}\ and\ \bibinfo {author} {\bibfnamefont {J.}~\bibnamefont
  {Redondo}},\ }\href {\doibase 10.1088/1475-7516/2012/02/032} {\bibfield
  {journal} {\bibinfo  {journal} {JCAP}\ }\textbf {\bibinfo {volume} {02}},\
  \bibinfo {pages} {032} (\bibinfo {year} {2012})},\ \Eprint
  {http://arxiv.org/abs/1110.2895} {arXiv:1110.2895 [hep-ph]} \BibitemShut
  {NoStop}%
\bibitem [{\citenamefont {Arias}\ \emph {et~al.}(2012)\citenamefont {Arias},
  \citenamefont {Cadamuro}, \citenamefont {Goodsell}, \citenamefont {Jaeckel},
  \citenamefont {Redondo},\ and\ \citenamefont {Ringwald}}]{Arias:2012az}%
  \BibitemOpen
  \bibfield  {author} {\bibinfo {author} {\bibfnamefont {P.}~\bibnamefont
  {Arias}}, \bibinfo {author} {\bibfnamefont {D.}~\bibnamefont {Cadamuro}},
  \bibinfo {author} {\bibfnamefont {M.}~\bibnamefont {Goodsell}}, \bibinfo
  {author} {\bibfnamefont {J.}~\bibnamefont {Jaeckel}}, \bibinfo {author}
  {\bibfnamefont {J.}~\bibnamefont {Redondo}}, \ and\ \bibinfo {author}
  {\bibfnamefont {A.}~\bibnamefont {Ringwald}},\ }\href {\doibase
  10.1088/1475-7516/2012/06/013} {\bibfield  {journal} {\bibinfo  {journal}
  {JCAP}\ }\textbf {\bibinfo {volume} {06}},\ \bibinfo {pages} {013} (\bibinfo
  {year} {2012})},\ \Eprint {http://arxiv.org/abs/1201.5902} {arXiv:1201.5902
  [hep-ph]} \BibitemShut {NoStop}%
\bibitem [{\citenamefont {Flacke}\ \emph {et~al.}(2017)\citenamefont {Flacke},
  \citenamefont {Frugiuele}, \citenamefont {Fuchs}, \citenamefont {Gupta},\
  and\ \citenamefont {Perez}}]{Flacke:2016szy}%
  \BibitemOpen
  \bibfield  {author} {\bibinfo {author} {\bibfnamefont {T.}~\bibnamefont
  {Flacke}}, \bibinfo {author} {\bibfnamefont {C.}~\bibnamefont {Frugiuele}},
  \bibinfo {author} {\bibfnamefont {E.}~\bibnamefont {Fuchs}}, \bibinfo
  {author} {\bibfnamefont {R.~S.}\ \bibnamefont {Gupta}}, \ and\ \bibinfo
  {author} {\bibfnamefont {G.}~\bibnamefont {Perez}},\ }\href {\doibase
  10.1007/JHEP06(2017)050} {\bibfield  {journal} {\bibinfo  {journal} {JHEP}\
  }\textbf {\bibinfo {volume} {06}},\ \bibinfo {pages} {050} (\bibinfo {year}
  {2017})},\ \Eprint {http://arxiv.org/abs/1610.02025} {arXiv:1610.02025
  [hep-ph]} \BibitemShut {NoStop}%
\bibitem [{\citenamefont {Gruber}\ \emph {et~al.}(1999)\citenamefont {Gruber},
  \citenamefont {Matteson}, \citenamefont {Peterson},\ and\ \citenamefont
  {Jung}}]{Gruber:1999yr}%
  \BibitemOpen
  \bibfield  {author} {\bibinfo {author} {\bibfnamefont {D.~E.}\ \bibnamefont
  {Gruber}}, \bibinfo {author} {\bibfnamefont {J.~L.}\ \bibnamefont
  {Matteson}}, \bibinfo {author} {\bibfnamefont {L.~E.}\ \bibnamefont
  {Peterson}}, \ and\ \bibinfo {author} {\bibfnamefont {G.~V.}\ \bibnamefont
  {Jung}},\ }\href {\doibase 10.1086/307450} {\bibfield  {journal} {\bibinfo
  {journal} {Astrophys. J.}\ }\textbf {\bibinfo {volume} {520}},\ \bibinfo
  {pages} {124} (\bibinfo {year} {1999})},\ \Eprint
  {http://arxiv.org/abs/astro-ph/9903492} {arXiv:astro-ph/9903492} \BibitemShut
  {NoStop}%
\bibitem [{\citenamefont {Bouchet}\ \emph {et~al.}(2008)\citenamefont
  {Bouchet}, \citenamefont {Jourdain}, \citenamefont {Roques}, \citenamefont
  {Strong}, \citenamefont {Diehl}, \citenamefont {Lebrun},\ and\ \citenamefont
  {Terrier}}]{Bouchet:2008rp}%
  \BibitemOpen
  \bibfield  {author} {\bibinfo {author} {\bibfnamefont {L.}~\bibnamefont
  {Bouchet}}, \bibinfo {author} {\bibfnamefont {E.}~\bibnamefont {Jourdain}},
  \bibinfo {author} {\bibfnamefont {J.~P.}\ \bibnamefont {Roques}}, \bibinfo
  {author} {\bibfnamefont {A.}~\bibnamefont {Strong}}, \bibinfo {author}
  {\bibfnamefont {R.}~\bibnamefont {Diehl}}, \bibinfo {author} {\bibfnamefont
  {F.}~\bibnamefont {Lebrun}}, \ and\ \bibinfo {author} {\bibfnamefont
  {R.}~\bibnamefont {Terrier}},\ }\href {\doibase 10.1086/529489} {\bibfield
  {journal} {\bibinfo  {journal} {Astrophys. J.}\ }\textbf {\bibinfo {volume}
  {679}},\ \bibinfo {pages} {1315} (\bibinfo {year} {2008})},\ \Eprint
  {http://arxiv.org/abs/0801.2086} {arXiv:0801.2086 [astro-ph]} \BibitemShut
  {NoStop}%
\bibitem [{\citenamefont {Essig}\ \emph {et~al.}(2013)\citenamefont {Essig},
  \citenamefont {Kuflik}, \citenamefont {McDermott}, \citenamefont {Volansky},\
  and\ \citenamefont {Zurek}}]{Essig:2013goa}%
  \BibitemOpen
  \bibfield  {author} {\bibinfo {author} {\bibfnamefont {R.}~\bibnamefont
  {Essig}}, \bibinfo {author} {\bibfnamefont {E.}~\bibnamefont {Kuflik}},
  \bibinfo {author} {\bibfnamefont {S.~D.}\ \bibnamefont {McDermott}}, \bibinfo
  {author} {\bibfnamefont {T.}~\bibnamefont {Volansky}}, \ and\ \bibinfo
  {author} {\bibfnamefont {K.~M.}\ \bibnamefont {Zurek}},\ }\href {\doibase
  10.1007/JHEP11(2013)193} {\bibfield  {journal} {\bibinfo  {journal} {JHEP}\
  }\textbf {\bibinfo {volume} {11}},\ \bibinfo {pages} {193} (\bibinfo {year}
  {2013})},\ \Eprint {http://arxiv.org/abs/1309.4091} {arXiv:1309.4091
  [hep-ph]} \BibitemShut {NoStop}%
\bibitem [{\citenamefont {Espinosa}\ \emph {et~al.}(1993)\citenamefont
  {Espinosa}, \citenamefont {Quiros},\ and\ \citenamefont
  {Zwirner}}]{Espinosa:1992kf}%
  \BibitemOpen
  \bibfield  {author} {\bibinfo {author} {\bibfnamefont {J.~R.}\ \bibnamefont
  {Espinosa}}, \bibinfo {author} {\bibfnamefont {M.}~\bibnamefont {Quiros}}, \
  and\ \bibinfo {author} {\bibfnamefont {F.}~\bibnamefont {Zwirner}},\ }\href
  {\doibase 10.1016/0370-2693(93)90450-V} {\bibfield  {journal} {\bibinfo
  {journal} {Phys. Lett. B}\ }\textbf {\bibinfo {volume} {314}},\ \bibinfo
  {pages} {206} (\bibinfo {year} {1993})},\ \Eprint
  {http://arxiv.org/abs/hep-ph/9212248} {arXiv:hep-ph/9212248} \BibitemShut
  {NoStop}%
\end{thebibliography}%

\end{document}